\documentclass[aps,reprint,twocolumn,superscriptaddress]{revtex4-1}
\usepackage{subfigure}
\usepackage{amsmath, amssymb,graphicx,color,bm,epstopdf}
\usepackage[dvipsnames]{xcolor}
\usepackage{bbold}
\usepackage{braket}
\usepackage{comment}
\usepackage{natbib}

\newcommand{\be}{\begin{equation}}
\newcommand{\e}{\end{equation}}
\newcommand{\beml}{\begin{subequations}}
\newcommand{\eml}{\end{subequations}}
\newcommand{\beq}{\begin{eqnarray}}
\newcommand{\eq}{\end{eqnarray}}
\newcommand{\ba}{\begin{array}}
\newcommand{\ea}{\end{array}}
\newcommand{\bpm}{\begin{pmatrix}}
\newcommand{\epm}{\end{pmatrix}}
\newcommand{\bc}{\begin{cases}}
\newcommand{\ec}{\end{cases}}

\newcommand{\0}{^{\phantom{\dagger}}}

\usepackage{placeins}
\usepackage{float}
\usepackage{upgreek}

\definecolor{amendments}{rgb}{0.0, 0.0, 0.7}

\usepackage[utf8]{inputenc}
\begin{document}
\title{Relativistic electron spin dynamics in a strong unipolar laser field}
\author{I.~A. Aleksandrov}
\affiliation{Department of Physics, St. Petersburg State University, 7/9 Universitetskaya Naberezhnaya, Saint Petersburg 199034, Russia}
\affiliation{Ioffe Institute, Politekhnicheskaya str. 26, Saint Petersburg 194021, Russia}
\author{D.~A. Tumakov}
\affiliation{Department of Physics, St. Petersburg State University, 7/9 Universitetskaya Naberezhnaya, Saint Petersburg 199034, Russia}
\author{A. Kudlis}
\affiliation{ITMO University, Kronverkskiy prospekt 49, Saint Petersburg 197101, Russia}
\author{V.~M. Shabaev}
\affiliation{Department of Physics, St. Petersburg State University, 7/9 Universitetskaya Naberezhnaya, Saint Petersburg 199034, Russia}
\author{N.~N. Rosanov}
\affiliation{Ioffe Institute, Politekhnicheskaya str. 26, Saint Petersburg 194021, Russia}
\affiliation{ITMO University, Kronverkskiy prospekt 49, Saint Petersburg 197101, Russia}

\begin{abstract}
The behavior of an electron spin interacting with a linearly polarized laser field is analyzed. In contrast to previous considerations of the problem, the initial state of the electron represents a localized wave packet, and a spatial envelope is introduced for the laser pulse, which allows one to take into account the finite size of both objects. Special attention is paid to ultrashort pulses possessing a high degree of unipolarity. Within a classical treatment (both nonrelativistic and relativistic), proportionality between the change of the electron spin projections and the electric field area of the pulse is clearly demonstrated. We also perform calculations of the electron spin dynamics according to the Dirac equation. Evolving the electron wave function in time, we compute the mean values of the spin operator in various forms. It is shown that the classical relativistic predictions are accurately reproduced when using the Foldy-Wouthuysen operator. The same results are obtained when using the Lorentz transformation and the nonrelativistic (Pauli) spin operator in the particle's rest frame.
\end{abstract}

\maketitle

\section{Introduction}

Investigations in the field of laser physics over many decades have not lost their relevance. Moreover, the appearance of new technological standards opens up broad prospects for understanding the complex processes that occur when matter interacts with a laser field (for a review see, e.g., Refs.~\cite{Sugioka2014,Phillips:15,Malinauskas2016,Sugioka2017}). In particular, the progress in designing laser setups which can generate femtosecond or even attosecond pulses~\cite{Brabec2000,Krausz2009,Keller2010,Wu2012,Manzoni2015,Ramasesha2016,Calegari2016,Hassan2016,You2017,Xu2018} motivates researchers to conduct more thorough investigations of various fundamental and practical problems. One of these issues is a comprehensive theoretical description of the interaction between ultrashort pulses and quantum objects. Possessing a number of remarkable features, among which, for instance, a high energy density, such laser pulses provide an effective tool for studying the atomic scale processes (see, e.g., Refs.~\cite{PhysRevLett.70.1236,Yogo2017,Reinhold_1995,Ramasesha2016,FUKUDA2007130}). Pulses with a high {\it degree of unipolarity}~\cite{archipov2017}, i.e., those whose electric field almost does not change its direction, are of particular importance here. The feasibility of generating such pulses was demonstrated in a number of studies~\cite{Kozlov2011,Arkhipov_2016,Pakhomov2017}. Quantitatively, a degree of unipolarity in the case of a spatially homogeneous field can be described by the following parameter:
\begin{equation}\label{UNIPOL}
    \upxi = \frac{|\int \! \boldsymbol{E}(t)dt|}{\int|\boldsymbol{E}(t)|dt},
\end{equation}
where $\boldsymbol{E}(t)$ is the corresponding electric field strength. The main advantage of pulses with large values of $\upxi$ is that they allow one to achieve maximum efficiency in problems related to acceleration of charged particles~\cite{Wu2012,Krausz2009}. The numerator in Eq.~\eqref{UNIPOL} represents the so-called \textit{electric field area of the pulse}~\cite{Rozanov2009},
\begin{equation}\label{sqEl}
\boldsymbol{S}_E = \int \!\! \boldsymbol{E}(t)dt. 
\end{equation}
If the field represents a finite laser pulse propagating in a certain direction, then one should integrate over $t$ for a given position in space, which gives essentially the area of the pulse profile. This quantity has several interesting properties, notably the fact that it remains constant when the electromagnetic pulse propagates through dissipative media (the properties of the electric field area of the pulse were discussed in detail in Refs.~\cite{Rozanov2009,Rozanov2010,Rozanov2015,2018rcharchpahroz,Rozanov2018,Rozanov2019}). It turns out that this parameter to a large extent determines the behavior of an electron in a laser field. For example, in Ref.~\cite{Rozanov2018_2} it was shown that the probability of hydrogen atom excitation can be approximately represented as a function depending solely on the electric area defined by Eq.~\eqref{sqEl}. In order to show it, an approximate solution of the nonstationary Schr\"odinger equation for a hydrogen atom in the presence of a laser field was found taking into account the specific properties of the considered pulses --- their ultrashort duration and high intensity. In the subsequent studies~\cite{Rozanov2019_2, rosanov_jetp_2020} considering a classical relativistic charged particle, it was also shown that in the case of the interaction of an accelerated particle with a laser pulse of arbitrary shape, the particle's final state is directly governed by the electric field area.

Along with studying the kinematic characteristics of an electron in external laser fields, the analysis of the dynamics of its intrinsic angular momentum --- spin --- is also of a great importance. Various aspects of this problem were addressed in a number of investigations (see Refs.~\cite{Walser_2000,Walser_2002,Peatross2007,Bauke_2014,Bauke_2014_2,Fu2019,Li2019}). For instance, in Ref.~\cite{Walser_2002} within a classical approach, the exact temporal dependence of the electron spin interacting with a plane monochromatic wave was obtained in both nonrelativistic and relativistic regimes. In addition, a nonrelativistic quantum-mechanical analysis of the problem was carried out. It was demonstrated that the electron spin precesses with a certain frequency around the magnetic field direction. However, the electron wave function was not localized in a major part of the previous investigations, nor was the spatial envelope introduced for the laser pulse within the scenario under consideration although studying the interaction between the two objects of a finite size should provide a solid connection to real experimental setups. For example, in Ref.~\cite{Walser_2000} the authors localized only the electron as a Gaussian wave packet keeping the external laser field infinite in space. In order to incorporate the spatiotemporal localization of the field, two different approaches are usually employed. The first one rests on the use of a temporal envelope which allowed one to smoothly turn on and off the external electromagnetic field (see, e.g., Refs.~\cite{Bauke_2014,Bauke_2014_2}. Such a treatment of the problem is basically required by the need for solving the Dirac equation within a finite time interval. However, a more natural approach to localizing the laser field is to introduce a spatial envelope making the field a finite pulse traveling along a certain direction. In this study, we follow the latter course describing it in detail in Sec.~\ref{sec:field}. Finally, we note that the quantum spin dynamics is often analyzed only on the basis of the Schr\"odinger equation, i.e., in the nonrelativistic framework (see, e.g., Ref.~\cite{Walser_2002}).

We aim to study the behavior of the electron spin interacting with a linearly polarized laser field within the classical formalism and the framework of relativistic quantum mechanics, where the initial electron state chosen in the form of a Gaussian wave packet evolves according to the Dirac equation. For the laser pulse, we introduce a spatial envelope in a similar way as was done recently in Ref.~\cite{Fu2019}, where, however, the spin was considered by means of a classical approach based on the Lagrangian formalism~\cite{Walser_1999,Barut_1990}, and the authors were primarily focused on the influence of the electron spin on its own kinematics (see also, e.g., Ref.~\cite{wen_pra_2017} and references therein). We place the main emphasis on studying the dynamics of the electron spin itself when interacting with laser pulses of a high degree of unipolarity. The present investigation is a natural continuation of a series of articles devoted to the analysis of the electron dynamics in ultrashort pulses~\cite{Rozanov2018_2,Rozanov2018_3,Rozanov2019_2}. Performing accurate calculations, we examine the role of the electric field area and compare the classical predictions with the results of our quantum simulations.

It is also important to note that the relativistic electron spin is well defined only in the absence of external fields exerting forces on the particle. Moreover, when the electron travels with a large velocity, the nonrelativistic (Pauli) operator $\hat{\boldsymbol{s}}_\text{P} = \boldsymbol{\Sigma}/2$ considered in the usual Dirac representation is no longer applicable. The problem of how one should describe {\it relativistic} spin effects remains highly contentious (see, e.g., Refs.~\cite{Bauke_2014_spin1, Bauke_2014_spin2,Caban2013, celeri_pra_2016, bliokh2017} and references therein). According to Refs.~\cite{Foldy1950, fradkin-good}, the quantum-mechanical counterpart of the classical spin vector is the Foldy-Wouthuysen operator~\cite{pryce_1948,Foldy1950}, i.e., the operator $\boldsymbol{\Sigma}/2$ considered within the Foldy-Wouthuysen representation (see also recent article~\cite{silenko_2020} where this issue is discussed in great detail). However, in the literature, there are numerous other operators that are considered as candidates for the spin operator (see, e.g., Refs.~\cite{Bauke_2014_spin1,Bauke_2014_spin2,Caban2013, celeri_pra_2016}). In the present study, we describe the electron spin dynamics by evaluating the mean values of the spin operator chosen in various forms. Besides the Pauli and Foldy-Wouthuysen operators, we will consider those of Frenkel~\cite{pryce_1948,hilgevoord1963,frenkel_1926,bmt_1959}, and Pryce~\cite{Bauke_2014_spin1,Bauke_2014_spin2,pryce_1948,stech,macfarlane}. As will be shown below, a very accurate agreement with the predictions of the classical relativistic model is achieved when using the Foldy-Wouthuysen operator, which is in accordance with the results of Refs.~\cite{Foldy1950, fradkin-good} (see also Ref.~\cite{silenko_2020} and references therein). Instead of Foldy-Wouthuysen operator, one can also employ the Pauli operator transformed from the particle's rest frame to the laboratory one. This operator is equivalent to the Foldy-Wouthuysen one within the subspace of the positive-energy states~\cite{chakrabarti1963, guersey1965, ryder1998, ryder1999}.

Our computations are based on the Dirac equation for an electron in the presence of a laser field in the form of a linearly polarized plane wave. To study the spin dynamics, we calculate the mean values of the spin projections on the Cartesian axes at the final time instant, when the electron and the laser pulse no longer interact. The exact wave function is constructed by means of the expansion coefficients with respect to the basis of the Volkov solutions~\cite{Wolkow1935}.

The paper is organized as follows. In Sec.~\ref{sec:field} we describe the field configuration of the laser pulse and the geometry of the process under consideration. In Sec.~\ref{sec:spin} we briefly discuss the choice of the relativistic electron spin operator. In Sec.~\ref{sec:plane_wave} we describe the method used for propagating the initial electron wave function. Sec.~\ref{sec:class} contains a classical analysis of the spin dynamics leading to approximate closed-form expressions for the final spin projections. In Sec.~\ref{sec:results} the main results of our numerical computations are presented and discussed. Finally, in Sec.~\ref{sec:conclusion} we draw a conclusion.

We use atomic units throughout the article: Planck constant $\hbar=1$, electron mass $m=1$, electron charge $e=-1$. In these units the speed of light in vacuum is $1/\alpha\approx 137.036$, where $\alpha$ is the fine structure constant.

\section{Description of the process}\label{sec:field}

\begin{figure*}[t]
\includegraphics[height=6.5cm]{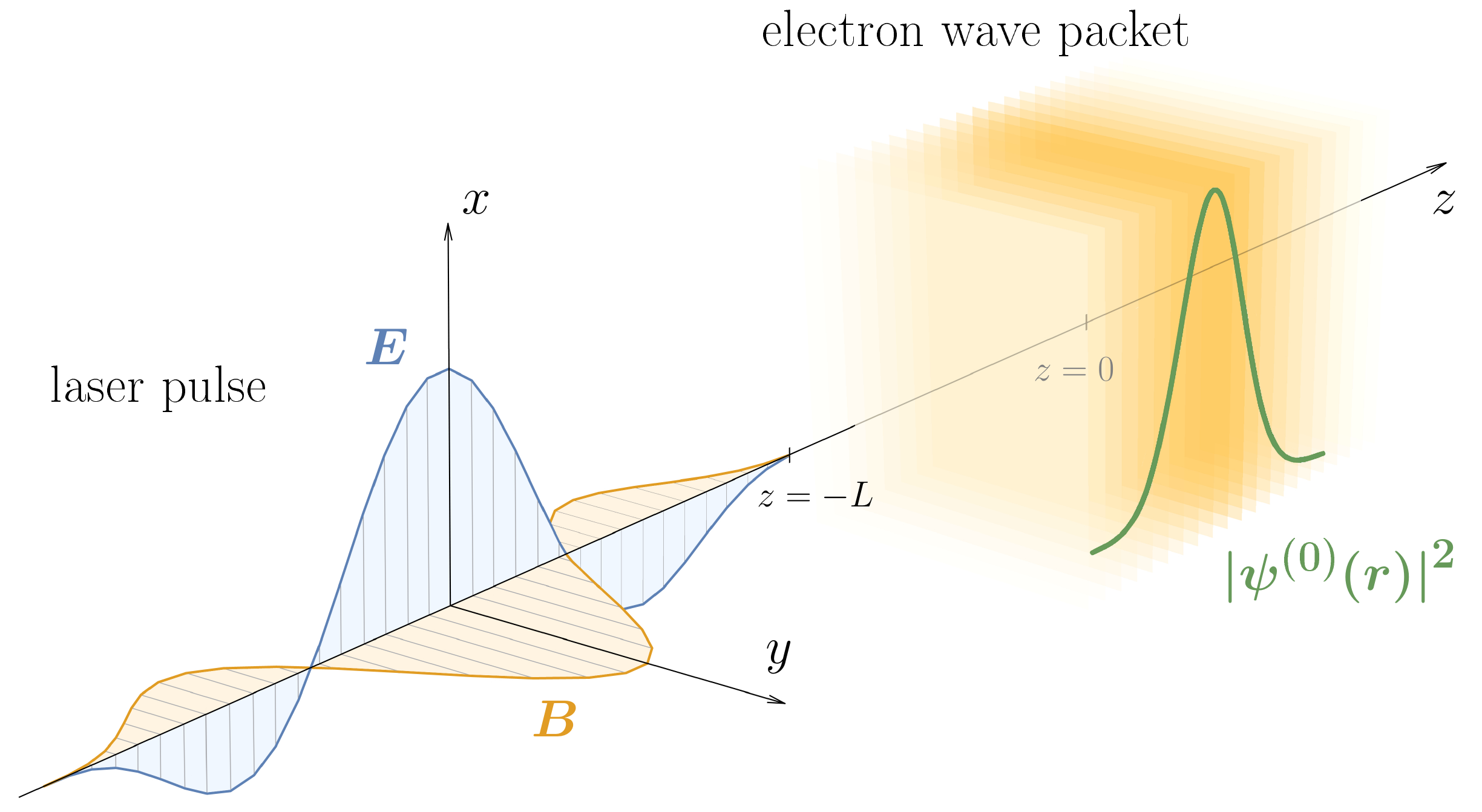}
\caption{Schematic setup of the process under consideration at $t = t_\text{in}$. The electron wave packet is localized along the $z$ axis and has initially a Gaussian profile with central value $z=0$. The laser pulse is polarized along the $x$ axis, the magnetic field $\boldsymbol{B}$ is directed along $y$, and the pulse travels in the $z$ direction. At $t = t_\text{in}$ the laser pulse and the wave packet do not overlap ($L$ is sufficiently large).}
\label{fig:scheme}
\end{figure*}

Both the laser pulse and the wave packet are assumed to be spatially localized only along the $z$ axis which coincides with the direction of the laser pulse propagation. The electromagnetic field is polarized along the $x$ axis and modeled with a vector potential in the form of a plane wave. Namely, the electric component of the field is chosen in the laboratory frame as follows:
\begin{eqnarray}
    E_x (t, z) &=& \mathcal{E}(ct - z),\label{eq:field_1}\\
    \mathcal{E}(\xi) &=& E_* F\left(\frac{\omega \xi}{c}\right) \sin \frac{\omega \xi}{c} ,
\label{eq:field_2}
\end{eqnarray}
where $E_*$ is the field amplitude, $\omega$ is the carrier frequency of the pulse, $c$ is the speed of light, and $t$ is time in the laboratory frame. Accordingly, only one component of the vector potential $A^\mu$ is not equal to zero:
\begin{equation}
A^1=A_x(t, z) = \mathcal{A}(ct - z).
\label{eq:vecPot}
\end{equation}
The vector $n^\mu=(1,0,0,1)$ satisfies the following relations: $n\cdot x\equiv n^\mu x_\mu = ct - z$ and $n^2 \equiv n^\mu n_\mu=0$, where $x^\mu = (ct, \boldsymbol{r})$. The corresponding wave vector is $\boldsymbol{k}_0 =(\omega/c)\boldsymbol{n}$. The function $F$ represents a smooth envelope which is chosen as
\begin{equation}
F(\eta) = \sin^2[\eta/(2N_\text{c})]\theta(\pi N_\text{c} - |\eta -\pi N_\text{c}|),
\end{equation}
so that the pulse contains $N_\text{c}$ carrier cycles (this number can be non-integer). In this study, we vary $N_\text{c}$ within the interval $0\leqslant N_\text{c} \leqslant 2$, where the degree of unipolarity $\upxi$ is large. Note that in the case of such small values of $N_\text{c}$, the carrier and the envelope of the pulse~\eqref{eq:field_2} cannot be evidently disentangled, nor should $\omega$ be interpreted as a well-defined fundamental frequency of the external field. Moreover, for small values of $N_\text{c}$, the field strength does not basically reach $E_*$. In what follows, we will consider $N_\text{c}$ as a parameter governing the pulse duration (for given $\omega$) and, more important, the electric field area of the pulse, which does not vanish once $N_\text{c}$ is non-integer.

The initial state of the setup is displayed in Fig.~\ref{fig:scheme}. At the initial time instant $t_\text{in} = -L/c$, the laser pulse is localized within the region $z\in[-L-\xi_\text{max}, \ -L]$, where $\xi_\text{max}\equiv 2\pi c N_\text{c}/\omega$. The value of $L$ should be large enough, so that the laser pulse and the electron wave packet do not overlap at $t = t_\text{in}$, i.e., the support of the wave function has to reside within the ray $z>-L$. The initial wave packet is centered at the origin $z=0$. The final state after the interaction is considered at $t=t_\text{out} = (\tilde{L} + \xi_\text{max})/c$, where $\tilde{L}$ is the position of the left edge of the laser pulse. It should be sufficiently large, so that the external field and the wave packet no longer overlap. We assume also that $\mathcal{A}=0$ for $\xi \leqslant 0$, while for $\xi \geqslant \xi_\text{max}$ it has an arbitrary value $\mathcal{A}_0$. The latter point allows us to examine a broad class of laser pulses including those having a large electric field area, e.g., unipolar ones.

Our main purpose is to calculate the mean values of the spin projections, which change due to the interaction of the electron with the external electromagnetic field. To this end, we construct the exact solution of the Dirac equation in the standard (Dirac) representation which includes the interaction with the laser pulse:
\begin{equation}
i \, \frac{\partial}{\partial t} \psi (t, \boldsymbol{r}) = \Big [ c \, \boldsymbol{\alpha} \cdot \Big \{ \hat{\boldsymbol{p}} + \frac{1}{c} \boldsymbol{A} (t, \boldsymbol{r}) \Big \} + \beta c^2 \Big ] \psi (t, \boldsymbol{r}).
\label{eq:dirac_equation}
\end{equation}
Here $\boldsymbol{A} (t, \boldsymbol{r}) = \mathcal{A} (ct - z) \boldsymbol{e}_x$, $\psi(t,\boldsymbol{r})$ is the electron wave function that coincides with the initial wave packet at $t=t_\text{in}$, $\hat{\boldsymbol{p}}=-i\boldsymbol{\nabla}$ is the momentum operator, $\boldsymbol{e}_x$ is the unit vector in the $x$ direction, and $\boldsymbol{\alpha}$ and $\beta$ are the Dirac matrices defined as follows:
\begin{equation}
\boldsymbol{\alpha} = \begin{pmatrix} 0 & \boldsymbol{\sigma} \\\boldsymbol{\sigma} & 0\end{pmatrix},  \quad \beta = \begin{pmatrix} I & 0 \\0 & -I \end{pmatrix},
\end{equation}
where $\boldsymbol{\sigma}$ are the Pauli matrices and $I$ is the identity matrix $2 \times 2$. Having constructed the exact wave function, we can then calculate the mean values of the spin projections:
\begin{equation}
\langle \hat{s}_i \rangle (t) = \int \! \psi^\dagger (t, \boldsymbol{r}) \hat{s}_i \psi (t, \boldsymbol{r}) d\boldsymbol{r},\quad i = 1,\, 2,\, 3.
\label{eq:spin_mean}
\end{equation}
The explicit form of the operator $\hat{\boldsymbol{s}}$ is not uniquely defined in relativistic quantum mechanics. This issue will be discussed in the next section. The method of constructing the wave function $\psi(t,\boldsymbol{r})$ is described in detail in Sec.~\ref{sec:plane_wave}.

\section{Relativistic spin operator} \label{sec:spin}

Even before the Dirac relativistic theory was formulated, Pauli proposed a quantum mechanical equation describing the motion of a charged particle with spin $1/2$ in an external electromagnetic field~\cite{pauli_1927}. This equation for a two-component wave function incorporates the energy of the interaction between the particle's intrinsic angular momentum (spin) and the magnetic field. The corresponding spin operator in the nonrelativistic theory has the form $\hat{\boldsymbol{s}}_\text{NR}=\boldsymbol{\sigma}/2$. The operator of the orbital angular momentum associated with the particle motion reads $\hat{\boldsymbol{l}} = \boldsymbol{r}\times\hat{\boldsymbol{p}}$. 
Both of these operators together with the total angular momentum operator $\hat{\boldsymbol{j}}_\text{NR}=\hat{\boldsymbol{l}}+\hat{\boldsymbol{s}}_\text{NR}$ commute with the nonrelativistic Hamiltonian in the absence of external fields, which means that all of the three vectors are conserved in the case of a free particle. A straightforward generalization of these expressions within relativistic quantum mechanics leads to the following:
\begin{equation}
\hat{\boldsymbol{s}}_\text{P} = \frac{1}{2} \boldsymbol{\Sigma},\quad \hat{\boldsymbol{j}} = \hat{\boldsymbol{l}} + \hat{\boldsymbol{s}}_\text{P}, \quad \boldsymbol{\Sigma} = \begin{pmatrix} \boldsymbol{\sigma} & 0 \\ 0 & \boldsymbol{\sigma} \end{pmatrix}.
\label{eq:spin_pauli}
\end{equation}
%
Nevertheless, there is no clear reason why these specific forms of the operators should be considered in the Dirac representation. The operator $\hat{\boldsymbol{s}}_\text{P}$, which acts in the space of four-component functions in the Dirac representation, will be referred to as the Pauli operator. In contrast to the nonrelativistic case, the operators $\hat{\boldsymbol{l}}$ and $\hat{\boldsymbol{s}}_\text{P}$ do not commute with the free-particle Dirac Hamiltonian $\hat{H}_\text{D}=c\,\boldsymbol{\alpha}\cdot\hat{\boldsymbol{p}} + \beta c^2$. It is only the total angular momentum $\boldsymbol{j}$ which represents a conserved quantity. However, as was shown in Refs.~\cite{Foldy1950, fradkin-good} (see also Ref.~\cite{silenko_2020}), the relativistic operator corresponding to the classical spin in the particle's rest frame is given by $\boldsymbol{\Sigma}/2$ in the {\it Foldy-Wouthuysen representation}. Let $\hat{U}_\text{FW}$ denote the Foldy-Wouthuysen unitary operator leading to two separate pairs of one-component equations which are equivalent to the four-component Dirac equation and independently describe the solutions with positive and negative energy, respectively~\cite{Foldy1950}. Then the Foldy-Wouthuysen spin operator $\hat{\boldsymbol{s}}_\text{FW}$ in the Dirac representation is the result of the transformation $\hat{U}^{-1}_\text{FW} \hat{\boldsymbol{s}}_\text{P} \hat{U}_\text{FW}$ which reads
\begin{equation}
\hat{\boldsymbol{s}}_\text{FW} = \frac{1}{2} \boldsymbol{\Sigma} + \frac{i\beta}{2\hat{p}_0} \, \hat{\boldsymbol{p}} \times \boldsymbol{\alpha} - \frac{\hat{\boldsymbol{p}} \times (\boldsymbol{\Sigma}\times \hat{\boldsymbol{p}})}{2\hat{p}_0 (\hat{p}_0 + c)},
\label{eq:spin_FW}
\end{equation}
where $\hat{p}_0 = \sqrt{c^2 + \hat{\boldsymbol{p}}^2}$. In the nonrelativistic limit, it obviously coincides with $\hat{\boldsymbol{s}}_\text{P} = \boldsymbol{\Sigma}/2$.

However, there are several other forms of the spin operator discussed in the literature besides the Pauli and Foldy-Wouthuysen ones (see Refs.~\cite{Bauke_2014_spin1,Bauke_2014_spin2,Caban2013,celeri_pra_2016}). In this investigation, we will also examine the operator in the form of Frenkel~\cite{pryce_1948,hilgevoord1963,frenkel_1926,bmt_1959}, which is defined by the following expression:
\begin{equation}
\hat{\boldsymbol{s}}_\text{F} = \frac{1}{2} \boldsymbol{\Sigma} + \frac{i\beta}{2c} \, \hat{\boldsymbol{p}} \times \boldsymbol{\alpha}.
\label{eq:spin_F}
\end{equation}
This operator can be obtained, for example, by applying Noether's theorem in the case of the Klein-Fock-Gordon theory formulated in the bispinor space~\cite{hilgevoord1963}. Both of the operators~\eqref{eq:spin_FW} and \eqref{eq:spin_F}, unlike $\hat{\boldsymbol{s}}_\text{P}$, commute with the Dirac Hamiltonian $\hat{H}_\text{D}$. Note, however, that the Frenkel operator does not satisfy the commutation relations $[\hat{s}_i,\,\hat{s}_j] = i\varepsilon_{ijk} \hat{s}_k$, and its eigenvalues are not equal to $\pm 1/2$. The definitions~\eqref{eq:spin_FW} and~\eqref{eq:spin_F} lead also to the following power expansion in $\hat{\boldsymbol{\pi}}\equiv\hat{\boldsymbol{p}}/c$:
\begin{eqnarray}
\hat{\boldsymbol{s}}_\text{FW} &=& \ \frac{1}{2} \boldsymbol{\Sigma} + \frac{i\beta}{2} (\hat{\boldsymbol{\pi}} \times \boldsymbol{\alpha}) \bigg (1 - \frac{1}{2} \hat{\boldsymbol{\pi}}^2 \bigg ) \nonumber \\ 
&-& \frac{1}{4} \hat{\boldsymbol{\pi}} \times (\boldsymbol{\Sigma} \times \hat{\boldsymbol{\pi}}) \bigg (1 - \frac{3}{4} \hat{\boldsymbol{\pi}}^2 \bigg ) + \mathcal{O} (\hat{\boldsymbol{\pi}}^5) \nonumber \\
&=& \ \hat{\boldsymbol{s}}_\text{F} + \mathcal{O} (\hat{\boldsymbol{\pi}}^2).
\label{eq:spin_expansion}
\end{eqnarray}
Consequently, the Frenkel operator is a sum of the nonrelativistic operator $\hat{\boldsymbol{s}}_\text{P}$ and the leading-order relativistic part of the Foldy-Wouthuysen operator. 

Finally, we also employ the so-called ``Pryce operator'' whose name was taken from Refs.~\cite{Bauke_2014_spin1,Bauke_2014_spin2} although it is not clear whether Pryce was the first to mention it. This operator was also considered in Refs.~\cite{stech, macfarlane} and has the following form:
\begin{equation}
\hat{\boldsymbol{s}}_\text{Pr} = \frac{1}{2} \beta \boldsymbol{\Sigma} + \frac{1}{2} (\boldsymbol{\Sigma} \cdot \hat{\boldsymbol{p}}) (1 - \beta) \, \frac{\hat{\boldsymbol{p}}}{\hat{\boldsymbol{p}}^2}.
\label{eq:spin_Pr}
\end{equation}
Note that for a given $c$-numbered vector $\boldsymbol{p}$, it does not depend on $|\boldsymbol{p}|$, so it can already be considered as a nonrelativistic operator which, however, does not match $\hat{\boldsymbol{s}}_\text{P}$ unlike all of the operators mentioned above. On the other hand, the spin projection onto the $\boldsymbol{p}$ axis (helicity) is exactly the same for $\hat{\boldsymbol{s}}_\text{P}$, $\hat{\boldsymbol{s}}_\text{FW}$, $\hat{\boldsymbol{s}}_\text{F}$, and $\hat{\boldsymbol{s}}_\text{Pr}$. Moreover, the Pryce operator commutes with $\hat{H}_\text{D}$, has the proper commutation relations, and has the eigenvalues $\pm 1/2$~\cite{Bauke_2014_spin1,Bauke_2014_spin2}. Besides Eqs.~\eqref{eq:spin_FW} and \eqref{eq:spin_Pr}, the Foldy-Wouthuysen and Pryce operators have other equivalent expressions (see, e.g., Ref.~\cite{Bauke_2014_2}). It turns out that the Pryce operator can be obtained from Eq.~\eqref{eq:spin_FW} if one replaces $\hat{p}_0$ with $\hat{H}_\text{D}/c$.

Given the presence of the external electromagnetic field, the momentum operator $\hat{\boldsymbol{p}}$ should be replaced with the sum $\hat{\boldsymbol{p}} + \boldsymbol{A}(t,\boldsymbol{r})/c$. The mean value of the Pauli operator $\hat{\boldsymbol{s}}_\text{P}$ is computed in the coordinate representation via Eq.~\eqref{eq:spin_mean}. To calculate the mean values of the operators~\eqref{eq:spin_FW}, \eqref{eq:spin_F},  and~\eqref{eq:spin_Pr}, we turn to the momentum representation, where each component of the operator $\hat{\boldsymbol{p}}$ is just a $c$ number. In the following section, we describe the method utilized in order to obtain the exact wave function $\psi(t,\boldsymbol{r})$.

We also note that in the presence of the external field, the Foldy-Wouthuysen operator does not precisely have the form~\eqref{eq:spin_FW} with $\hat{\boldsymbol{p}} \to \hat{\boldsymbol{p}} + \boldsymbol{A}(t,\boldsymbol{r})/c$ as the corresponding unitary operator $\hat{U}_\text{FW}$ becomes less trivial. However, since we are interested in computing the {\it total} change of the spin projections, we consider the final electron state when the particle no longer interacts with the laser pulse, so the expression~\eqref{eq:spin_FW} is exact, provided one properly takes into account a nonzero (but constant) value of the vector potential in the space-time region where the final electron wave packet is localized. Furthermore, as the final state of the electron is free, the Foldy-Wouthuysen and Pryce operators yield exactly the same results. Indeed, in the absence of external fields, one can define the energy and momentum, so the operators $\hat{H}_\text{D}$ and $c \hat{p}_0$ are equivalent. We performed our calculations using both of Eqs.~\eqref{eq:spin_FW} and \eqref{eq:spin_Pr} and confirmed this point numerically. Accordingly, the results obtained by means of the Pryce operator will not be presented in what follows. Note that in the presence of external fields, these two operators are not equivalent.

Finally, one may also argue that instead of using some specific form of the spin operator, the spin degree of freedom is to be described within the particle's rest frame, which can easily be attained by performing the Lorentz boost from the laboratory frame if there are no external forces. One can then calculate the mean value of a certain projection of $\hat{\boldsymbol{s}}_\text{P}$ taking into account that the Lorentz transformation does not preserve the norm (it is not unitary). It turns out that within the subspace of the positive-energy states this approach is completely equivalent to the use of the Foldy-Wouthuysen operator and the operator $\hat{L}_{\boldsymbol{p}} \hat{\boldsymbol{s}}_\text{P} \hat{L}^{-1}_{\boldsymbol{p}}$, where $\hat{L}_{\boldsymbol{p}}$ is the corresponding Lorentz boost from the electron rest frame~\cite{chakrabarti1963, guersey1965, ryder1998, ryder1999}, i.e., the positive-energy eigenvectors of $\hat{\boldsymbol{s}}_\text{P}$ transformed by $\hat{L}_{\boldsymbol{p}}$ are the eigenvectors of $\hat{\boldsymbol{s}}_\text{FW}$ and $\hat{L}_{\boldsymbol{p}} \hat{\boldsymbol{s}}_\text{P} \hat{L}^{-1}_{\boldsymbol{p}}$. Accordingly, the use of the Lorentz transformations would lead to exactly the same findings regarding the problem considered in the present study. We also point out that in Refs.~\cite{Bauke_2014_spin1,Bauke_2014_spin2,Caban2013,celeri_pra_2016} a number of other different forms of the relativistic spin operator were examined.

\section{Calculation of the electron wave packet dynamics} 
\label{sec:plane_wave}

Due to the fact that the external field does not depend
on the coordinates $x$ and $y$, the corresponding components of the generalized momentum of the electron are conserved. To study the nontrivial dynamics of the electron with regard to the $z$ axis, we construct the initial wave packet as follows:
\begin{equation}
\psi^\text{(0)}_{\boldsymbol{p}, s} (\boldsymbol{r}) = \frac{1}{(2 \pi)^{3/2}} \, \mathrm{e}^{i \boldsymbol{p} \boldsymbol{r}} \! \int \limits_{-\infty}^{+\infty} \! dq \, \mathrm{e}^{iqz} f(q) u (\boldsymbol{p} + q \boldsymbol{n}, s),
\label{eq:psi_init}
\end{equation}
where $f(q)$ determines the spectral structure of the wave packet, $s$ is the spin quantum number, and $\boldsymbol{n}$ coincides with the unit vector $\boldsymbol{e}_z$. The vector $\boldsymbol{p}=(p_x,p_y,p_z)$ consists of the following components: $p_x$ and $p_y$ are the exact values of the projections of the electron momentum along the $x$ and $y$ axes, respectively, $p_z$ is the mean value of the $z$ projection. In order to ensure the condition $\langle \psi^\text{(0)}_{\boldsymbol{p}, s} | \psi^\text{(0)}_{\boldsymbol{p}, s'} \rangle = \delta_{s,s'}$, we require
\begin{equation}
\int \limits_{-\infty}^{+\infty} \! dq |f(q)|^2 = 1.
\end{equation}
The function $f(q)$ is chosen in the Gaussian form:
\begin{equation}
    f(q) = \frac{1}{\sqrt{\Delta q \sqrt{\pi}}} \, \mathrm{e}^{- q^2 / (2 \Delta q^2)},
\end{equation}
where the parameter $\Delta q$ governs the width of the wave packet. The initial spin state of the electron is determined by the constant (independent of coordinates and time) bispinors $u(\boldsymbol{p}, s)$ corresponding to the positive-energy solutions of the Dirac equation ($s =\pm$). Together with the bispinors $v(\boldsymbol{p},s)$ involved in the states with negative energy, they form a complete orthonormal set:
\begin{eqnarray}
u^\dagger (\boldsymbol{p}, s) u (\boldsymbol{p},s') = v^\dagger (\boldsymbol{p},s) v (\boldsymbol{p},s') &=& \delta_{ss'},\\
u^\dagger (\boldsymbol{p},s) v (\boldsymbol{p},s') &=& 0, \\ \sum_{s=\pm 1} \big [ u (\boldsymbol{p},s) u^\dagger (\boldsymbol{p},s) + v (\boldsymbol{p},s) v^\dagger (\boldsymbol{p},s) \big ] &=& I.
\end{eqnarray}
These bispinors satisfy the relations
\begin{eqnarray}
\big ( c \, \boldsymbol{\alpha} \cdot \boldsymbol{p} + \beta c^2 \big ) u (\boldsymbol{p},s) &=& \varepsilon u (\boldsymbol{p},s), \label{eq:de_u}\\
\big ( c \, \boldsymbol{\alpha} \cdot \boldsymbol{p} + \beta c^2 \big ) v (\boldsymbol{p},s) &=& -\varepsilon v (\boldsymbol{p},s), \label{eq:de_v}
\end{eqnarray}
where $\varepsilon = c \, \sqrt{c^2 + \boldsymbol{p}^2}$. We choose the bispinors in the following form:
\begin{eqnarray}
u (\boldsymbol{p}, +1) &=& \frac{1}{2\sqrt{p^0 (p^0 - p_x)}}
\begin{pmatrix}
c+p^0- p_x+i p_y\\
 p_z\\
 p_z\\
c-p^0+ p_x+i p_y
\end{pmatrix},\label{eq:ups_x_p}\\
u (\boldsymbol{p}, -1) &=& \frac{1}{2\sqrt{p^0 (p^0 - p_x)}}
\begin{pmatrix}
- p_z\\
c+p^0- p_x-i p_y\\
c-p^0+ p_x-i p_y\\
- p_z
\end{pmatrix},\label{eq:ups_x_m}\\
v (\boldsymbol{p}, +1) &=& \frac{1}{2\sqrt{p^0 (p^0 + p_x)}}
\begin{pmatrix}
c-p^0- p_x+i p_y\\
 p_z\\
 p_z\\
c+p^0+ p_x+i p_y
\end{pmatrix},\label{eq:vps_x_p} \\ 
v (\boldsymbol{p}, -1) &=& \frac{1}{2\sqrt{p^0 (p^0 + p_x)}}
\begin{pmatrix}
- p_z\\
c-p^0- p_x-i p_y\\
c+p^0+ p_x-i p_y\\
- p_z
\end{pmatrix},\label{eq:vps_x_m}
\end{eqnarray}
where $p^0=\varepsilon/c$. Note that for these bispinors the value of the quantum number $s$ corresponds to a certain spin projection ($\pm1/2$) onto the $z$ axis only in the case $p_z= 0$ (no matter which spin operator is employed). 

The initial condition reads $\psi_{\boldsymbol{p}, s} (t_\text{in}, \boldsymbol{r}) = \psi^\text{(0)}_{\boldsymbol{p}, s} (\boldsymbol{r})$. Our goal is to evolve this state in time and calculate the mean values of the spin projections. The main idea of the method is the following. The initial state can be expanded into the complete set of the Volkov states~\cite{Wolkow1935,boca2010,dipiazza2018}, and the expansion coefficients do not depend on time since the Dirac Hamiltonian is Hermitian. The wave function at an arbitrary time instant $t$ can then be constructed using the coefficients evaluated. The Volkov states are defined by the following expressions:
%
\begin{widetext}
\begin{eqnarray}
\varphi^{(\zeta)}_{\boldsymbol{p}', s'} (t, \boldsymbol{r}) &=& \frac{1}{(2 \pi)^{3/2}} \, \mathrm{e}^{i \zeta \boldsymbol{p}' \boldsymbol{r}} f^{(\zeta)}_{\boldsymbol{p}', s'} (t, z),\label{eq:volkov_1}\\
f^{(\zeta)}_{\boldsymbol{p}', s'} (t, z) &=& \mathrm{e}^{- i \zeta \varepsilon' t}  \, \mathrm{exp} \Bigg \{ - i \int \limits_0^{n \cdot x} \! d\xi \, \frac{1}{2 (n \cdot p')} \left[-\frac{2}{c} \, (p' \cdot A(\xi)) - \zeta \, \frac{1}{c^2} \, A^2(\xi) \right] \Bigg \} \nonumber \\
{}&\times &\bigg [ 1 - \frac{\zeta}{2c (n \cdot p')} \, (\gamma \cdot n) (\gamma \cdot A) \bigg ] w_\zeta (\boldsymbol{p}', s'),\label{eq:volkov_2}
\end{eqnarray}
\end{widetext}
where $w_+ (\boldsymbol{p}', s') = u (\boldsymbol{p}', s')$ and $w_- (\boldsymbol{p}', s') = v (-\boldsymbol{p}', s')$. Each of the Volkov functions has a well-defined sign of energy $\zeta=\pm$, which does not depend on time (this is consistent with the fact that a plane-wave electromagnetic field cannot produce electron-positron pairs). Given the specific form of the vector potential~\eqref{eq:vecPot} used in this paper, we obtain
%
\begin{widetext}
\begin{eqnarray}
f^{(\zeta)}_{\boldsymbol{p}', s'} (t, z) &=& \mathrm{e}^{- i \zeta \varepsilon' t}  \, \mathrm{exp} \Bigg \{ \frac{(-i)}{\varepsilon' - c p_z'} \Bigg [ p_x' \int \limits_0^{\xi} \! d\xi' \, \mathcal{A} (\xi')+ \frac{\zeta}{2c} \int \limits_0^{\xi} \! d\xi' \, \mathcal{A}^2 (\xi') \Bigg ] \Bigg \} \notag \\
{} &\times&  \bigg [ 1 + \frac{\zeta}{2 (\varepsilon' - c p_z')} \, \mathcal{A} (\xi) (\gamma^0 - \gamma^3) \gamma^1 \bigg ] w_\zeta (\boldsymbol{p}', s'),\label{eq:f_volkov}
\end{eqnarray}
\end{widetext}
where $\xi = n\cdot x = ct - z$.

The electron wave function can be expanded in terms of the Volkov states:
\begin{equation}
\psi_{\boldsymbol{p}, s} (t, \boldsymbol{r}) = \sum_{\zeta} \sum_{s'} \int \! d\boldsymbol{p}' \, C^{(\zeta)}_{\boldsymbol{p}', s'} \varphi^{(\zeta)}_{\boldsymbol{p}', s'} (t, \boldsymbol{r}).
\label{eq:expansion_psi_volkov}
\end{equation}
The expansion coefficients $C^{(\zeta)}_{\boldsymbol{p}',s'}$ are evaluated at $t=t_\text{in}$ as a standard inner product,
\begin{equation}
C^{(\zeta)}_{\boldsymbol{p}', s'} = \int \! d\boldsymbol{r} \, \big [\varphi^{(\zeta)}_{\boldsymbol{p}', s'} (t_\text{in}, \boldsymbol{r}) \big ]^\dagger \psi^\text{(0)}_{\boldsymbol{p}, s} (\boldsymbol{r}).
\end{equation}
As the wave packet~\eqref{eq:psi_init} depends on $x$ and $y$ only via $\mathrm{exp}(i \boldsymbol{p} \boldsymbol{r})$, the coefficients are ``diagonal'' with respect to $p_x$ and $p_y$:
\begin{equation}
C^{(\zeta)}_{\boldsymbol{p}', s'} = \delta (p_x' - \zeta p_x) \delta (p_y' - \zeta p_y) c^{(\zeta)}_{p_z', s'}.
\end{equation}
One can easily verify that
\begin{equation}
-i \partial_x \varphi^{(+)}_{\boldsymbol{p}', s'} (t, \boldsymbol{r}) = p_x' \varphi^{(+)}_{\boldsymbol{p}', s'} (t, \boldsymbol{r}).
\end{equation}
Thus the index $p_x'$ corresponds to the generalized momentum projection (the same holds also for $p_y'$). We receive
\begin{eqnarray}
c^{(\zeta)}_{p_z', s'} &=& \int \limits_{-\infty}^{+\infty} \! \frac{dz}{2 \pi} \int \limits_{-\infty}^{+\infty} \! dq \, \mathrm{e}^{i (p_z - \zeta p_z')z} \mathrm{e}^{iqz} 
\nonumber\\
&\times& \big [ f^{(\zeta)}_{\boldsymbol{p}', s'} (t_\text{in}, z) \big ]^\dagger f(q) u (\boldsymbol{p} + q \boldsymbol{n}, s),
\label{eq:coeff}
\end{eqnarray}
where $p'_x=\zeta p_x$ and $p_y' = \zeta p_y$. Since the initial state~\eqref{eq:psi_init} is orthogonal to the subspace of the negative-energy states, the coefficients $c^{(-)}_{\boldsymbol{p}',s'}$ vanish, which allows us to use only the Volkov solutions corresponding to positive energy ($\zeta = +$). The wave function can now be obtained according to
\begin{equation}
\psi_{\boldsymbol{p}, s} (t, \boldsymbol{r}) = \sum_{s'} \int \!  dp_z' \, c^{(+)}_{p_z', s'} \, \varphi^{(+)}_{p_x, p_y, p_z', s'} (t, \boldsymbol{r}).
\label{eq:psi_c_t}
\end{equation}
We use this expression at $t = t_\text{out}$ in order to evaluate the full change of the spin projections. Once the coefficients~\eqref{eq:coeff} are calculated, we build a spatial grid within a box whose center coincides with the classical value of the final coordinate $z$ (classical equations of motion are solved as a usual Cauchy problem). Then we adjust the box position and size to properly capture the final wave packet and calculate the wave function according to Eq.~\eqref{eq:psi_c_t} with necessary precision and spatial resolution. The same procedure is used for the box in momentum space.

When the exact wave function~(\ref{eq:psi_c_t}) is constructed, one can calculate the mean values of various observable quantities, e.g., the spin projections, either in the momentum or coordinate representation.

\section{Spin dynamics of a classical electron}
\label{sec:class}

The temporal dependence of the spin angular momentum of a classical electron in the presence of an external electromagnetic field can be described in the classic nonrelativistic case by means of the precession equation for the magnetic moment $\boldsymbol{m} = - \boldsymbol{s}/c$~($|\boldsymbol{s}|= 1/2$)~\cite{jacksonED}:
\begin{equation}
    \frac{d \boldsymbol{s}}{dt} = -\frac{1}{c} \boldsymbol{s} \times \left( \boldsymbol{B} - \frac{\boldsymbol{v}}{c} \times \boldsymbol{E} \right),
    \label{eq:larmor}
\end{equation}
where $\boldsymbol{v}$ is the electron's velocity. As was shown in Ref.~\cite{Walser_2002}, if the external field represents a monochromatic plane wave, i.e.,
\begin{equation}
\boldsymbol{E}=E_*\cos(\omega t - kz) \boldsymbol{e}_x,~~\boldsymbol{B}=E_*\cos(\omega t - kz) \boldsymbol{e}_y,
\label{eq:field_mono}
\end{equation}
where $k=\omega/c$, then the spin projections change according to
\begin{eqnarray}
    \Delta s_x^{\rm NR}(\tau) &=& \sin [\sigma_E (\tau)] \cos[\theta_0 + \sigma_E (\tau)], \label{eq:class-spin_x}\\
    \Delta s_y^{\rm NR} (\tau) &=& 0, \label{eq:class-spin_y}\\
    \Delta s_z^{\rm NR} (\tau) &=& -\sin [\sigma_E (\tau)] \sin[\theta_0 + \sigma_E (\tau)]. \label{eq:class-spin_z}
\end{eqnarray}
Here $\tau = t - z/c$, $\sigma_E (\tau) = S_E(\tau)/(2c)$, $S_E(\tau) = (E_*/\omega) \sin \omega \tau$ is the $x$ projection of the electric field area of the pulse calculated over a finite time interval, and $\theta_0$ determines the initial orientation of the particle's spin (in contrast to the notations of Ref.~\cite{Walser_2002}, $\theta_0$ is measured here from the $z$ direction). If the electric field area is sufficiently small, i.e., $|\sigma_E (\tau)| \ll 1$, one obtains
\begin{widetext}
\begin{eqnarray}
    \Delta s_x^{\rm NR \; approx.} (\tau) &= & \sigma_E (\tau) \cos \theta_0 - \sigma^2_E (\tau) \sin \theta_0, \label{eq:class-spin_x_nr} \\
    \Delta s_z^{\rm NR \; approx.} (\tau) & = & - \sigma_E (\tau) \sin \theta_0 - \sigma^2_E (\tau) \cos \theta_0, \label{eq:class-spin_z_nr}
\end{eqnarray}
\end{widetext}
where we have neglected the terms of order $\sigma^3_E (\tau)$ and higher. If $\theta_0 = 0$, the changes of the $x$ and $z$ spin projections are proportional to $\sigma_E (\tau)$ and $\sigma^2_E (\tau)$, respectively (see also Ref.~\cite{Peatross2007}).

A relativistic generalization of Eq.~\eqref{eq:larmor} is the Thomas-Bargmann-Michel-Telegdi (T-BMT) equation~\cite{bmt_1959, thomas_1927} (see also, e.g., Refs.~\cite{jacksonED, mane_2005, frenkel_1926}),
\begin{equation}
    \frac{d \boldsymbol{s}}{dt} = -\frac{1}{c} \boldsymbol{s} \times \left(\frac{1}{\gamma} \boldsymbol{B} - \frac{1}{\gamma+1} \frac{\boldsymbol{v}}{c} \times \boldsymbol{E} \right),
    \label{eq:thomas}
\end{equation}
where $\gamma=(1 - \boldsymbol{v}^2/c^2)^{-1/2}$. In the case of the monochromatic field~\eqref{eq:field_mono}, one can derive the relativistic analogues of the relations~\eqref{eq:class-spin_x}--\eqref{eq:class-spin_z} (see Ref.~\cite{Walser_2002}):
\begin{eqnarray}
    \Delta s_x^{\rm R}(\tau) &=& \sin [\arctan \{\sigma_E (\tau)\}] \nonumber \\
    &\times& \cos [ \theta_0 + \arctan \{\sigma_E (\tau)\}], \label{eq:rel-spin_x}\\
    \Delta s_y^{\rm R} (\tau) &=& 0, \label{eq:rel-spin_y}\\ 
    \Delta s_z^{\rm R} (\tau) &=& -\sin [\arctan \{\sigma_E (\tau)\}] \nonumber \\
    &\times& \sin [\theta_0 + \arctan \{\sigma_E (\tau)\}] \label{eq:rel-spin_z}.
\end{eqnarray}
In the case $\sigma_E (\tau) \ll 1$, one recovers the expressions \eqref{eq:class-spin_x}--\eqref{eq:class-spin_z_nr}. In Ref.~\cite{Walser_2002} these results were obtained assuming that the particle is initially at rest. In what follows, we will also consider a nonzero initial momentum $p_z$. In this case, Eqs.~\eqref{eq:rel-spin_x}--\eqref{eq:rel-spin_z} alter according to the substitution $\arctan \{\sigma_E (\tau)\} \to \arctan \{ \sigma_E (\tau)/D \}$, where $D \equiv (1 + \Pi_z - p_z/c)/2$ and $\Pi_z \equiv [1+(p_z/c)^2]^{1/2}$. This modification is always taken into account in our computations and is important unless $|p_z/c| \ll 1$ as $D = 1 - p_z/(2c) +[p_z/(2c)]^2 + \mathcal{O}(|p_z/c|^4)$. The derivation of this result can be found in Appendix. We also note that Eqs.~\eqref{eq:rel-spin_x}--\eqref{eq:rel-spin_z} have this particular form in terms of $\sigma_E (\tau)$ no matter what phase is chosen in Eq.~\eqref{eq:field_mono} (one can, for instance, replace $\cos$ with $\sin$).

In addition to using the analytical expressions~\eqref{eq:class-spin_x}--\eqref{eq:class-spin_z_nr} and \eqref{eq:rel-spin_x}--\eqref{eq:rel-spin_z}, we also solved numerically the equations of motion for a particle in the field of a finite laser pulse~\eqref{eq:field_1}--\eqref{eq:field_2} and evolved the spin angular momentum according to Eqs.~\eqref{eq:larmor} and \eqref{eq:thomas}. Thus, in our calculations only quantum effects were not taken into account. In the next section, we compare these predictions for the case of a classical electron with the analytical expressions for the case of a monochromatic field and with the results of quantum calculations described in Sec.~\ref{sec:plane_wave}. In order to partially take into account the finite size of the laser pulse when studying the total change of 
the spin, we replace the area $S_E(\tau)$ in Eqs.~\eqref{eq:class-spin_x}--\eqref{eq:class-spin_z_nr} and \eqref{eq:rel-spin_x}--\eqref{eq:rel-spin_z} with the total electric field area of the laser field~\eqref{eq:field_1}--\eqref{eq:field_2},
\begin{equation}
S_E = \begin{cases} (E_*/\omega) \sin^2 (\pi N_\text{c})/(1 - N_\text{c}^2), & N_\text{c} \neq 1, \\ 0, & N_\text{c} = 1. \end{cases}
\label{eq:S_E_Nc}
\end{equation}
As will be seen below, this substitution to a great extent takes into account the effects of the spatial finiteness of the laser pulse.

Finally, we consider $p_z = 0$ and $\theta_0 = 0$, which substantially simplifies the expressions displayed above, so that they take the following form:
\begin{eqnarray}
    \Delta s_x^{\rm NR}(\tau) &=& \frac{1}{2} \sin [2\sigma_E (\tau)], \label{eq:dsx_nr_theta0}\\
    \Delta s_x^{\rm NR \; approx.}(\tau) &=& \sigma_E (\tau), \label{eq:dsx_nr_app_theta0}\\
    \Delta s_x^{\rm R}(\tau) &=& \frac{\sigma_E (\tau)}{1+\sigma^2_E (\tau)}, \label{eq:dsx_rel_theta0}\\
   \Delta s_z^{\rm NR}(\tau) &=& - \sin^2 [\sigma_E (\tau)], \label{eq:dsz_nr_theta0}\\
    \Delta s_z^{\rm NR \; approx.}(\tau) &=& -\sigma^2_E (\tau), \label{eq:dsz_nr_app_theta0}\\
    \Delta s_z^{\rm R}(\tau) &=& -\frac{\sigma^2_E (\tau)}{1+\sigma^2_E (\tau)}. \label{eq:dsz_rel_theta0}
\end{eqnarray}
These classical relations clearly demonstrate that the electron spin dynamics is fully determined by the electric field area of the external laser field, which indicates a great impact which unipolar pulses have on the particle's spin. In the next section, these approximate expressions and those for $\theta_0$, $p_z \neq 0$ [Eqs.~\eqref{eq:class-spin_x}--\eqref{eq:class-spin_z_nr} and Eqs.~\eqref{eq:rel-spin_x}--\eqref{eq:rel-spin_z}] will be benchmarked against the numerical solutions of Eqs.~\eqref{eq:larmor} and \eqref{eq:thomas} and the results of quantum simulations.

\section{Results and discussion}
\label{sec:results}

In this section, we discuss the predictions of the classical treatment of the electron spin dynamics and the results of relativistic calculations based on the Dirac equation which is solved by means of the method presented in Sec.~\ref{sec:plane_wave}. The mean value of the electron spin is evaluated using the spin operators discussed in Sec.~\ref{sec:spin}. Our computations are carried out for various values of the particle's initial central momentum $p_z$ (the transverse components of the momentum are equal to zero). We choose first the following external field parameters: $E_* = 10~\text{a.u.} \approx 0.514~\text{V/cm}$ and $\omega=1$~a.u. corresponding to a peak intensity of $3.51\times 10^{18}$~W/cm${}^2$ and the wavelength $\lambda = 2\pi c/\omega \approx 45.6$~nm. According to Eq.~\eqref{eq:S_E_Nc}, the total electric field area always satisfies $|S_E| \lesssim 14~\text{a.u.} \approx 1.74 \times 10^{-4}~\text{V}\cdot \text{s/m}$, so $|\sigma_E| \lesssim 0.05$. In section~\ref{sec:res_low_freq}, we will also examine laser pulses with $\omega = 0.1$~a.u. ($\lambda \approx 456$~nm), which can have a large electric field area, i.e., $|\sigma_E| \ll 1$ no longer holds for such pulses.

The width $\Delta q$ of the initial electron wave packet in momentum space was varied from $0.0001$~a.u. to $1$~a.u. in our quantum computations, which, in fact, did not affect the results presented in what follows. Moreover, the final mean values of the $z$ coordinate calculated with the operator $\boldsymbol{r}$ and the $z$ projection of the particle's momentum proved to exactly follow the classical relativistic solutions, i.e., the corresponding relative discrepancy was always much smaller than the spin effects examined in this study. It means that the spin-induced forces, which we do not incorporate in our classical treatment, are insignificant within our simulations. We also found that the difference between the results obtained with the operator $\boldsymbol{r}$ in the Dirac and Foldy-Wouthuysen representations is negligible.

Finally, we point out that we entirely neglect the QED effects and radiation reaction since the external field strength is not large enough to manifest them (see, e.g., Refs.~\cite{esirkepov_pla_2015, del_sorbo_2018, blackburn_2020}).
\begin{figure}[t]
\includegraphics[width=0.48\textwidth]{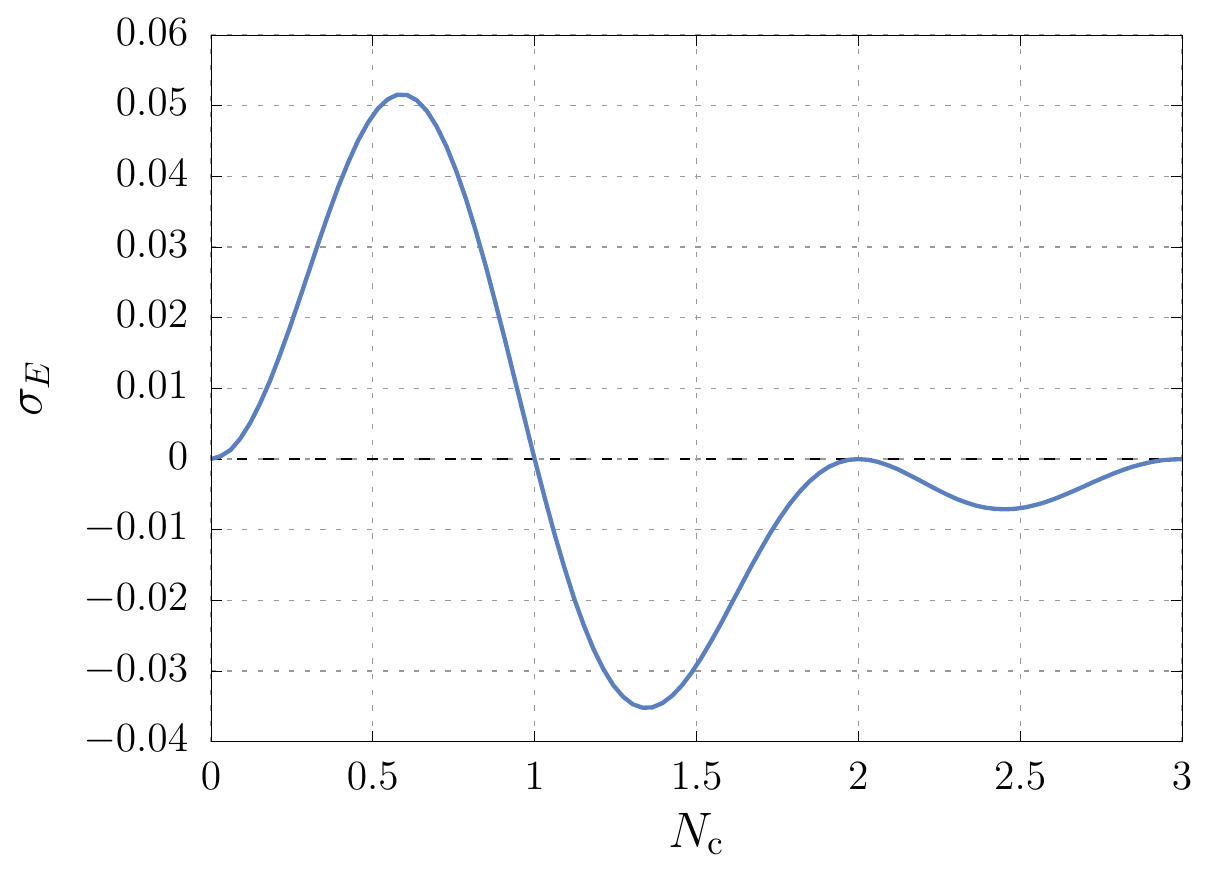}
\caption{Total dimensionless electric field area $\sigma_E = S_E/(2c)$ as a function of $N_\text{c}$ evaluated according to Eq.~\eqref{eq:S_E_Nc} for $E_* = 10$~a.u. and $\omega=1$~a.u.}
\label{fig:area}
\end{figure}
\begin{figure*}[t]
\subfigure{\label{fig:sx_class_14}\includegraphics[height=6.1cm]{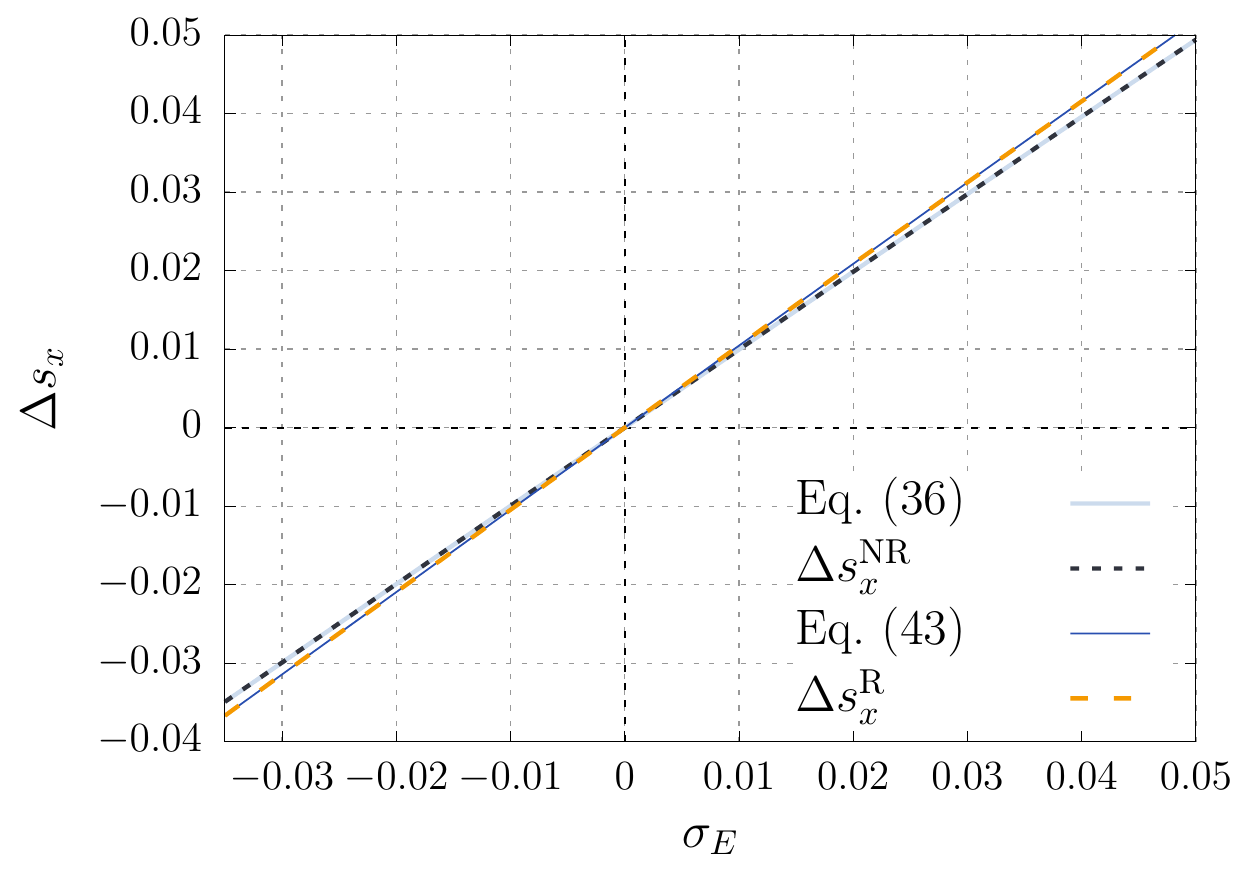}}~~~\subfigure{\label{fig:sz_class_14}\includegraphics[height=6.1cm]{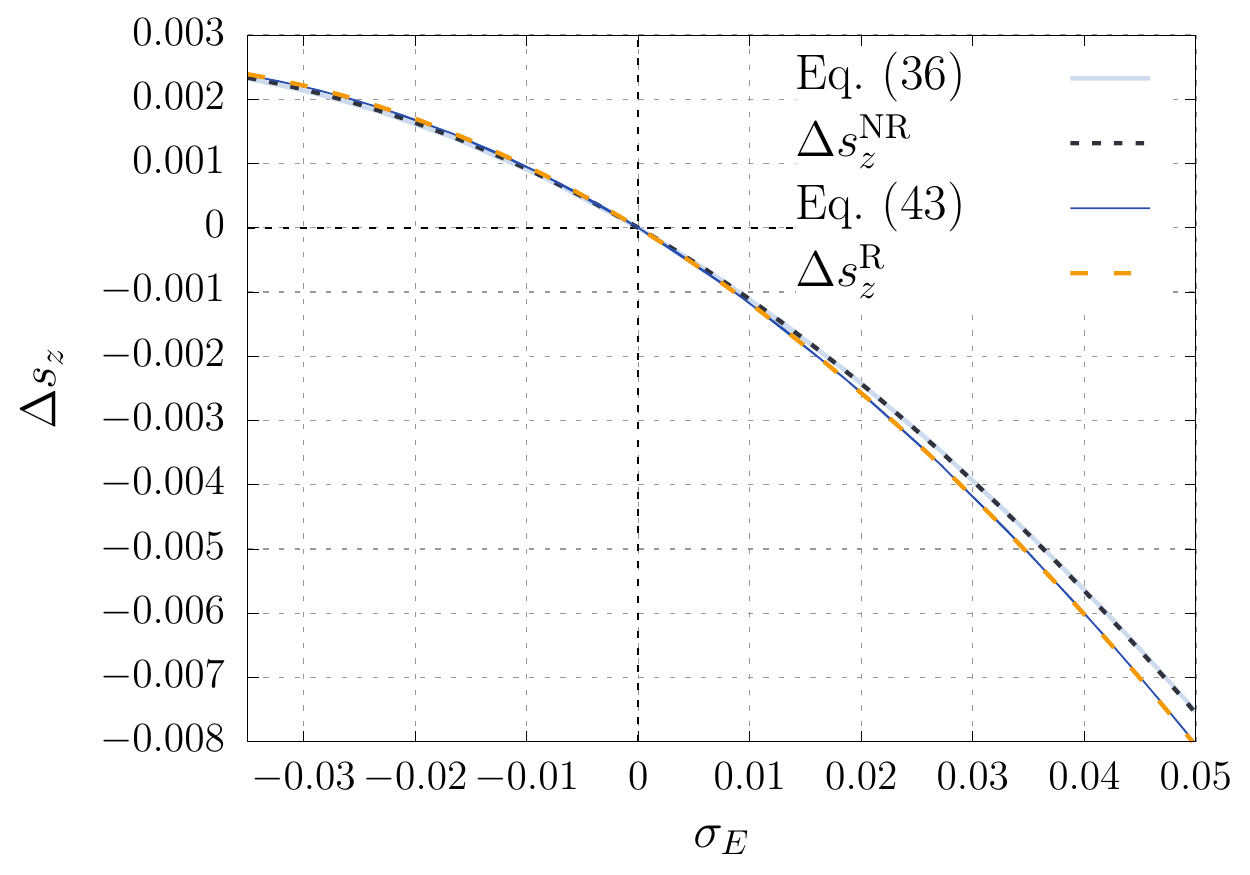}}
\caption{Change of the spin projections $s_x$~(left) and $s_z$~(right) of a classical electron after the interaction with the laser pulse~\eqref{eq:field_1}--\eqref{eq:field_2} as a function of the electric field area. The approximate predictions $\Delta s^\text{NR}$ and $\Delta s^\text{R}$ are obtained with the aid of Eqs.~\eqref{eq:class-spin_x}, \eqref{eq:class-spin_z} and Eqs.~\eqref{eq:rel-spin_x}, \eqref{eq:rel-spin_z}, respectively, using the actual electric field area~\eqref{eq:S_E_Nc}. The equations~\eqref{eq:larmor} and \eqref{eq:thomas} are solved numerically taking into account the spatiotemporal dependence of the laser field~\eqref{eq:field_1}--\eqref{eq:field_2}. The initial momentum of the electron is $p_z=14$~a.u. ($p_x = p_y = 0$), and $\theta_0 = 0.102 \approx 6^\circ$. The external field parameters are $E_* = 10$~a.u., $\omega = 1$~a.u.}
\label{fig:s_class_14}
\end{figure*}
\begin{figure*}[t]
\subfigure{\label{fig:sx_class_70}\includegraphics[height=6.1cm]{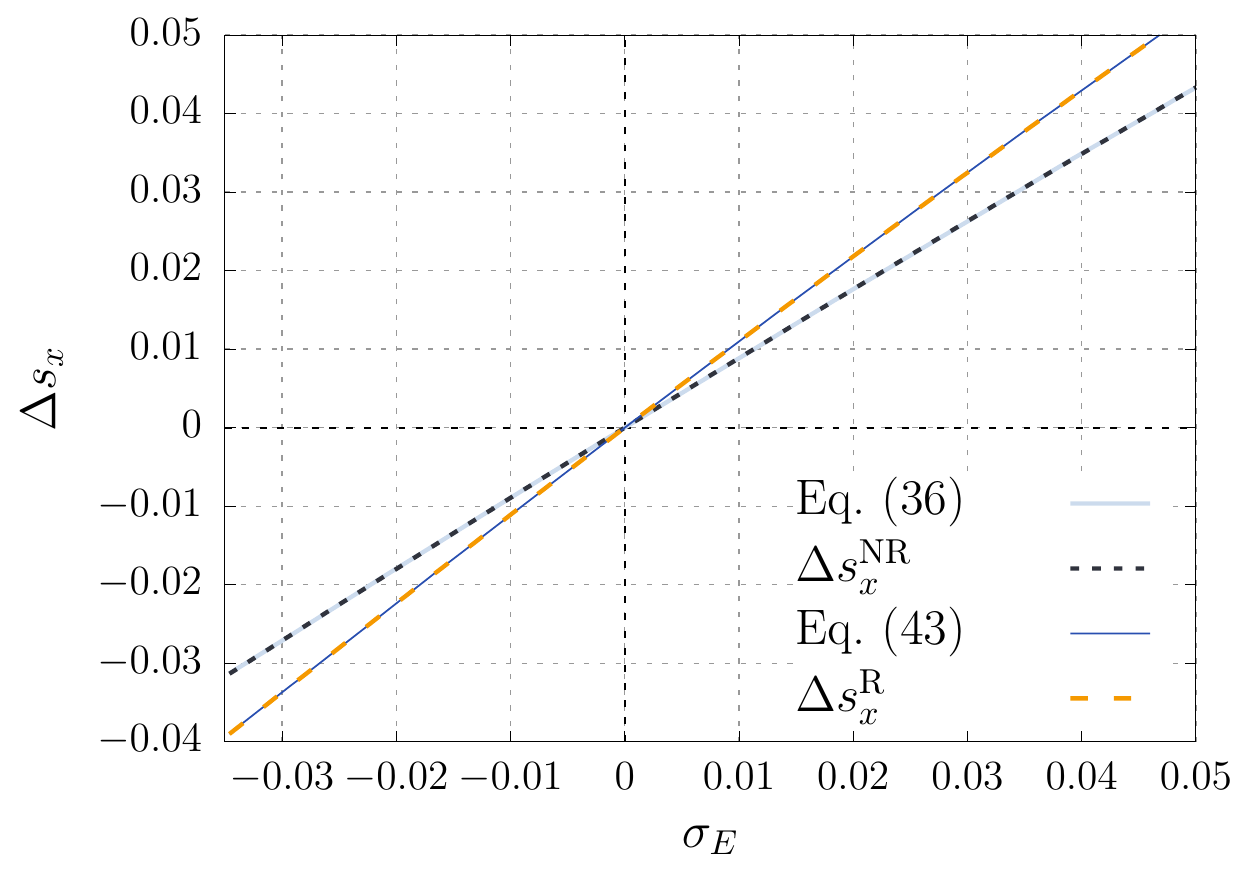}}~~~\subfigure{\label{fig:sz_class_70}\includegraphics[height=6.1cm]{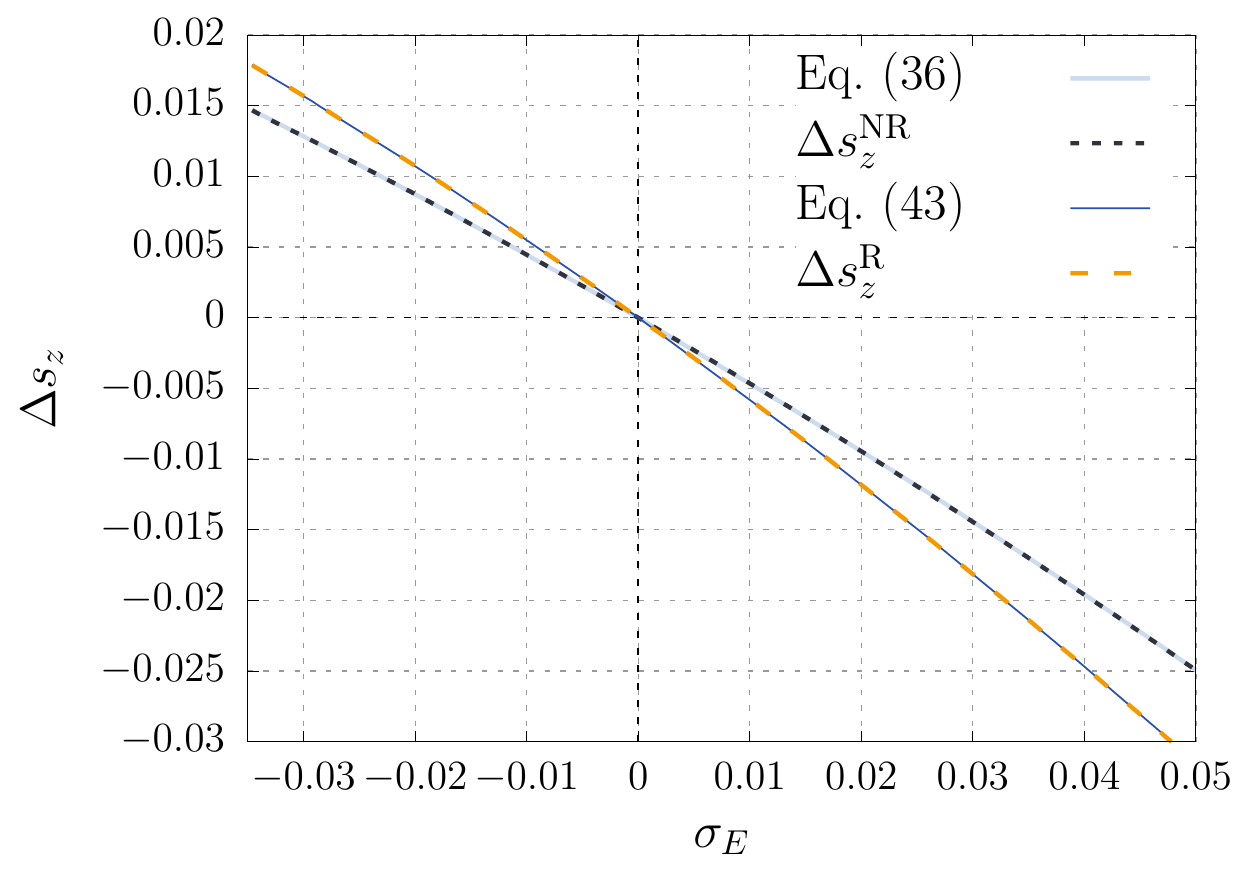}}
\caption{Change of the spin projections $s_x$~(left) and $s_z$~(right) of a classical electron after the interaction with the laser pulse~\eqref{eq:field_1}--\eqref{eq:field_2} as a function of the electric field area. The plot legend is the same as in Fig.~\ref{fig:s_class_14}. The initial momentum of the electron is $p_z=70$~a.u. ($p_x = p_y = 0$), and $\theta_0 = 0.472 \approx 27^\circ$.}
\label{fig:s_class_70}
\end{figure*}

\subsection{Classical spin dynamics} \label{sec:res_class}

Let us first consider the classical treatment of the problem where we compare the predictions of Eqs.~\eqref{eq:class-spin_x}--\eqref{eq:class-spin_z_nr} and Eqs.~\eqref{eq:rel-spin_x}--\eqref{eq:rel-spin_z} with the results of the exact calculations based on the classical equations~\eqref{eq:larmor} and \eqref{eq:thomas}. To elucidate the influence of the laser field on the electron spin, we evaluate the total change of the spin projections as a difference between their final and initial values.

First, we note that in all our calculations, the change of the $y$ projection of the spin (projection onto the magnetic field direction) was always at least 5 orders of magnitude smaller than the corresponding values for the projections along the $x$ and $z$ axes, which agrees well with the results~\eqref{eq:class-spin_y} and~\eqref{eq:rel-spin_y}. Accordingly, we will refrain from discussing the $y$ spin projection and will only analyze the $x$ and $z$ ones. Second, since the electric field area is sufficiently small, the approximate expressions~\eqref{eq:class-spin_x_nr} and \eqref{eq:class-spin_z_nr} yield the same results as those obtained by means of Eqs.~\eqref{eq:class-spin_x} and \eqref{eq:class-spin_z} (the corresponding lines would be indistinguishable from one another in the plots presented in this subsection).

Before we discuss how the total spin change depends on the dimensionless field area $\sigma_E = S_E/(2c)$, we present the plot $\sigma_E (N_\text{c})$ according to Eq.~\eqref{eq:S_E_Nc} which is used throughout the paper (see Fig.~\ref{fig:area}). This function oscillates and vanishes for integer values of $N_\text{c}$. The amplitude decreases with $N_\text{c}$. As we are mainly interested in laser pulses which have a high degree of unipolarity~\eqref{UNIPOL}, we opt to consider only pulses containing no more than two ``optical cycles'' ($0<N_\text{c}\leqslant 2$).

In Fig.~\ref{fig:s_class_14} we display the total change of the electron spin projections as a function of the field area $\sigma_E$ for $p_z = 14$~a.u. ($p_x=p_y=0$). Using the expression~\eqref{eq:S_E_Nc} for the total electric field area of the laser pulse~\eqref{eq:field_1}--\eqref{eq:field_2} allows us to partially take into consideration the finiteness of the laser pulse in Eqs.~\eqref{eq:class-spin_x}, \eqref{eq:class-spin_z}, \eqref{eq:rel-spin_x}, and \eqref{eq:rel-spin_z}, which were derived in the case of a monochromatic field~\eqref{eq:field_mono}. In order to take into account the finite-size effects precisely, we performed the exact numerical computations evolving the classical particle's spin according to Eqs.~\eqref{eq:larmor} and \eqref{eq:thomas} (solid lines in Fig.~\ref{fig:s_class_14}).

The plots in Fig.~\ref{fig:s_class_14} uncover several important patterns. First, we observe that plugging the actual electric field area into the approximate expressions allows one to capture the effects of the spatial finiteness of the external laser pulse to very high accuracy, i.e., the exact solutions of Eqs.~\eqref{eq:larmor} and \eqref{eq:thomas} yield the same results. It means also that the change of the electron spin is governed by very simple closed-form expressions, e.g., Eqs.~\eqref{eq:rel-spin_x}--\eqref{eq:rel-spin_z} in the relativistic regime, and it is determined by the electric field area of the laser pulse, which plays a crucial role in the process. Second, the shape of the curves in Fig.~\ref{fig:s_class_14}(left) differs from that of the curves in Fig.~\ref{fig:s_class_14}(right). This can be easily accounted for by means of Eqs.~\eqref{eq:class-spin_x_nr}--\eqref{eq:class-spin_z_nr}. The initial value $\theta_0$ of the precession angle amounts to $\theta_0 = 0.102 \approx 6^\circ$, which matches the corresponding expectation value for the Foldy-Wouthuysen spin operator in the initial electron state within our quantum simulations [the bispinors~\eqref{eq:ups_x_p}--\eqref{eq:vps_x_m} correspond to nonzero $\theta_0$ once $p_z \neq 0$]. Since both $\theta_0$ and $\sigma_E$ are small, the right-hand side of Eq.~\eqref{eq:class-spin_x_nr} is almost linear in $\sigma_E$, which leads to the straight lines in Fig.~\ref{fig:s_class_14}(left). On the other hand, both of the terms in Eq.~\eqref{eq:class-spin_z_nr} are significant (they both are considerably smaller than $\Delta s^\text{NR}_x$). Thus, in Fig.~\ref{fig:s_class_14}(right) one observes a parabola whose vertex corresponds to $\sigma_E^* = -(1/2) \tan \theta_0 \approx -0.051$. Finally, the graphs reveal a discrepancy between the nonrelativistic and relativistic predictions which is expected to grow with increasing $p_z$. We note that for $p_z \to 0$ all of the four curves plotted completely coincide.

In Fig.~\ref{fig:s_class_70} we depict the results of the analogous calculations with $p_z = 70$~a.u. ($p_z \sim c/2$). In this case, $\theta_0 = 0.472$. First, we see that the curves in Fig.~\ref{fig:s_class_70}(right) are now much closer to straight lines due to a large value of $\theta_0$ ($\approx 27^\circ$) making the first (linear) term in Eq.~\eqref{eq:class-spin_z_nr} dominant. Observe also that the change $\Delta s_z$ becomes notably larger. Besides, the discrepancy between the relativistic and nonrelativistic results is now well pronounced, as it should be (such great values of $p_z/c$ do not warrant using nonrelativistic methods). On the other hand, the approximate treatment of the finite-size effects remains very accurate as there is no difference between the solid and dashed lines in the graphs.

Our results indicate that the change of the particle's spin strongly depends on the electric field area of the laser pulse and can be described to high precision by the approximate formulas~\eqref{eq:class-spin_x}, \eqref{eq:class-spin_z}, \eqref{eq:rel-spin_x}, and \eqref{eq:rel-spin_z}, provided one employs a proper form of the function $\sigma_E$ depending on the external field parameters. The classical analysis suggests that unipolar laser pulses are particularly efficient at changing the electron spin state. Finally, we note that the field area $S_E$ (or dimensionless $\sigma_E$) is the relevant quantity here unlike the unipolarity parameter~$\upxi$. 

\begin{figure*}[t]
\subfigure{\label{fig:sx_t_0}\includegraphics[height=6.1cm]{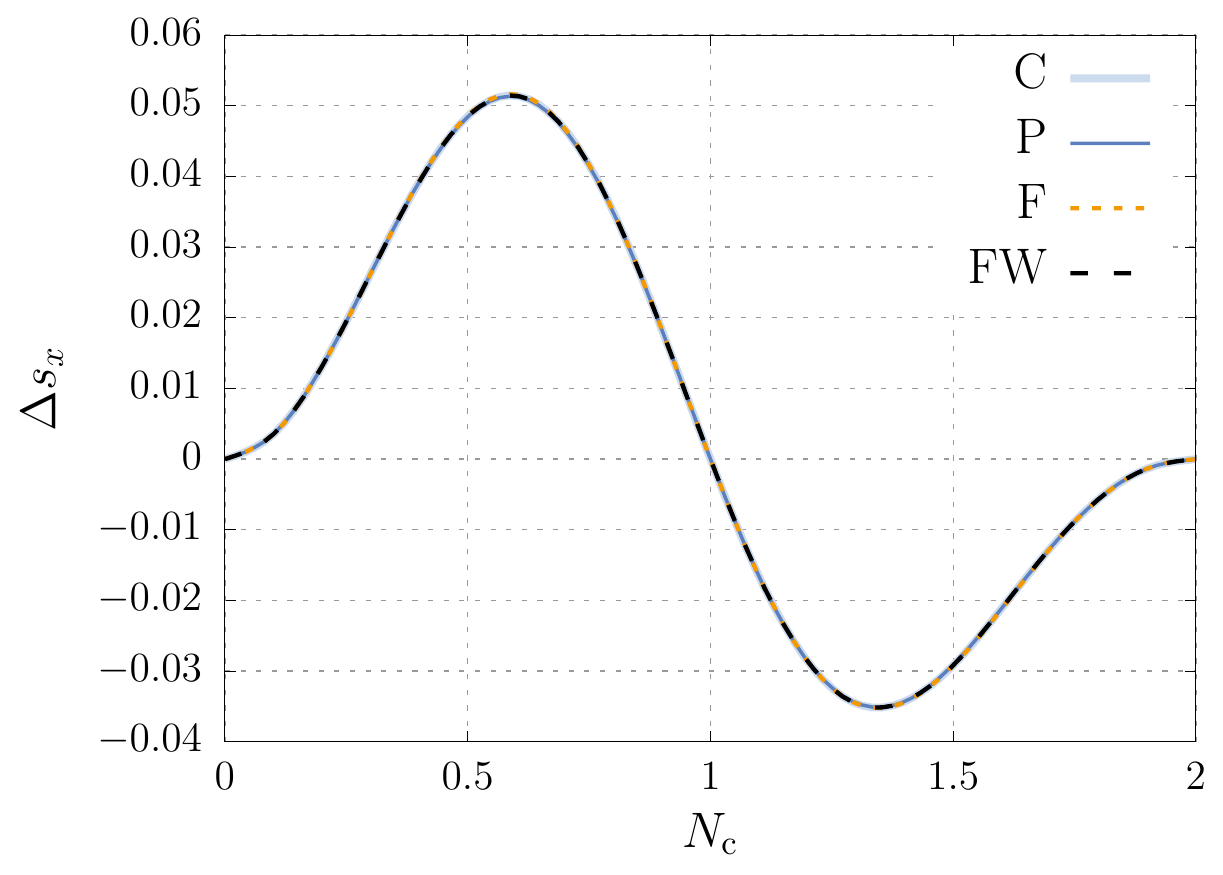}}~~~\subfigure{\label{fig:sz_t_0}\includegraphics[height=6.1cm]{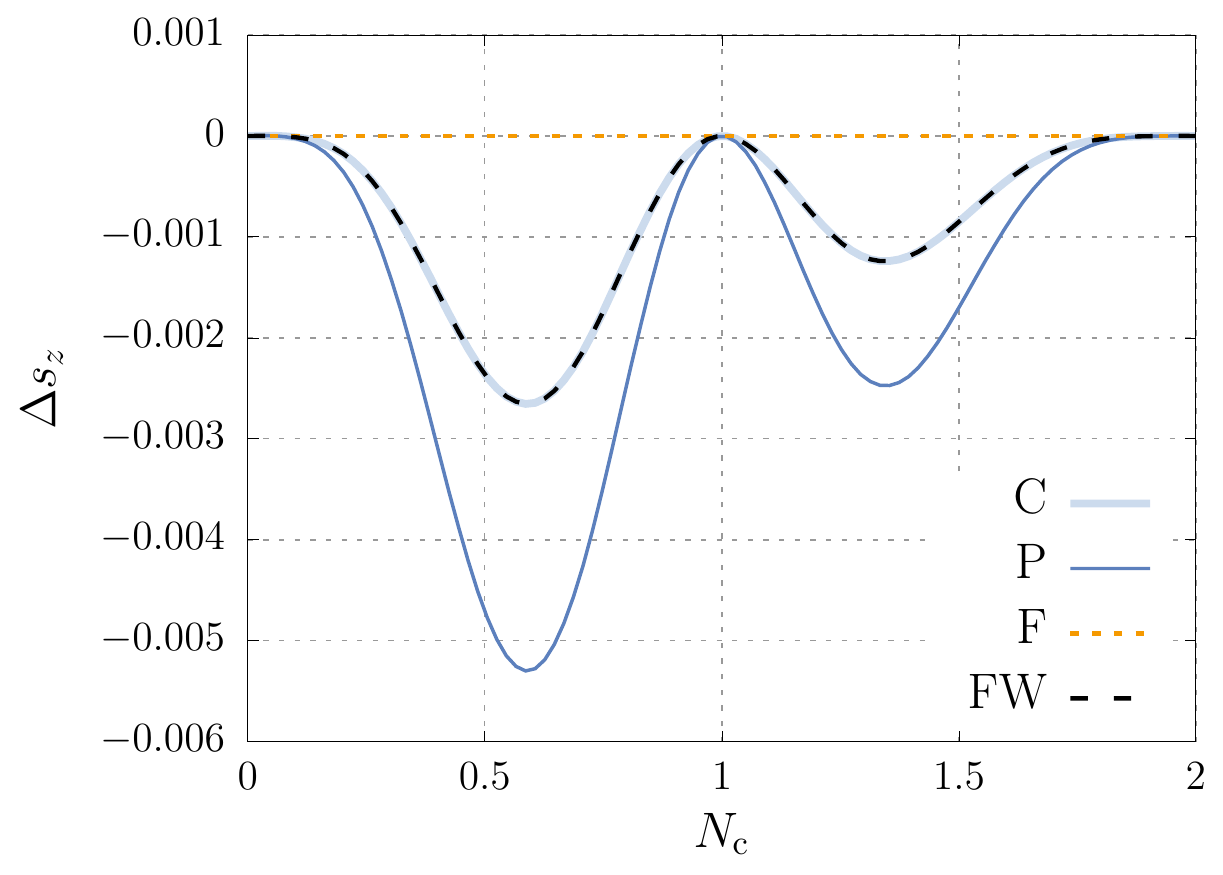}}
\caption{Change of the spin projections $s_x$~(left) and $s_z$~(right) of a relativistic electron wave packet after the interaction with the laser pulse~\eqref{eq:field_1}--\eqref{eq:field_2} as a function of $N_\text{c}$. The calculations are performed with the aid of the classical T-BMT equation~\eqref{eq:thomas} (line ``C'') and by solving the Dirac equation and using the Pauli, Frenkel, and Foldy-Wouthuysen spin operators (lines ``P'', ``F'', and ``FW'', respectively). The initial electron momentum is $\boldsymbol{p}=0$. The external field parameters are $E_* = 10$~a.u., $\omega = 1$~a.u.}
\label{fig:s_t_0}
\end{figure*}
\begin{figure*}
\subfigure{\label{fig:sx_t_14}\includegraphics[height=6.1cm]{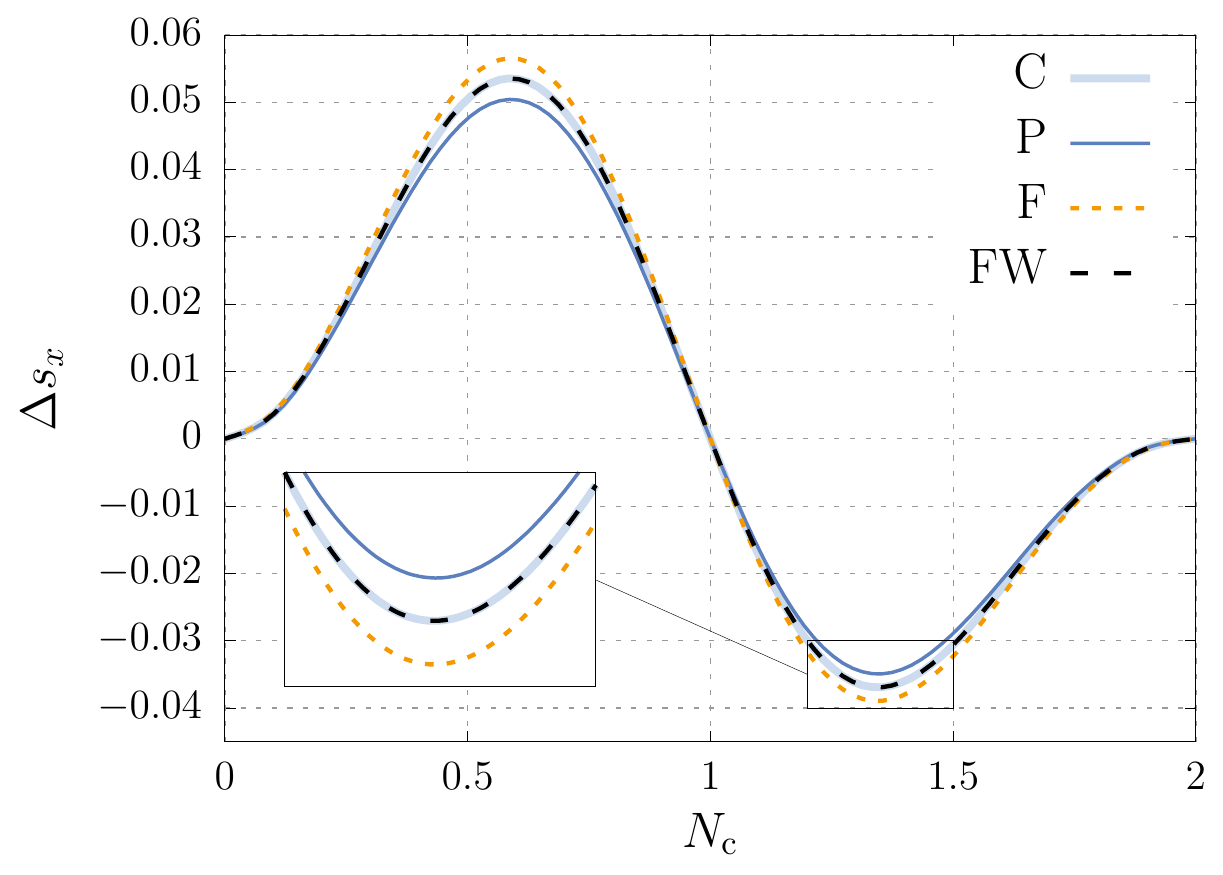}}~~~\subfigure{\label{fig:sz_t_14}\includegraphics[height=6.1cm]{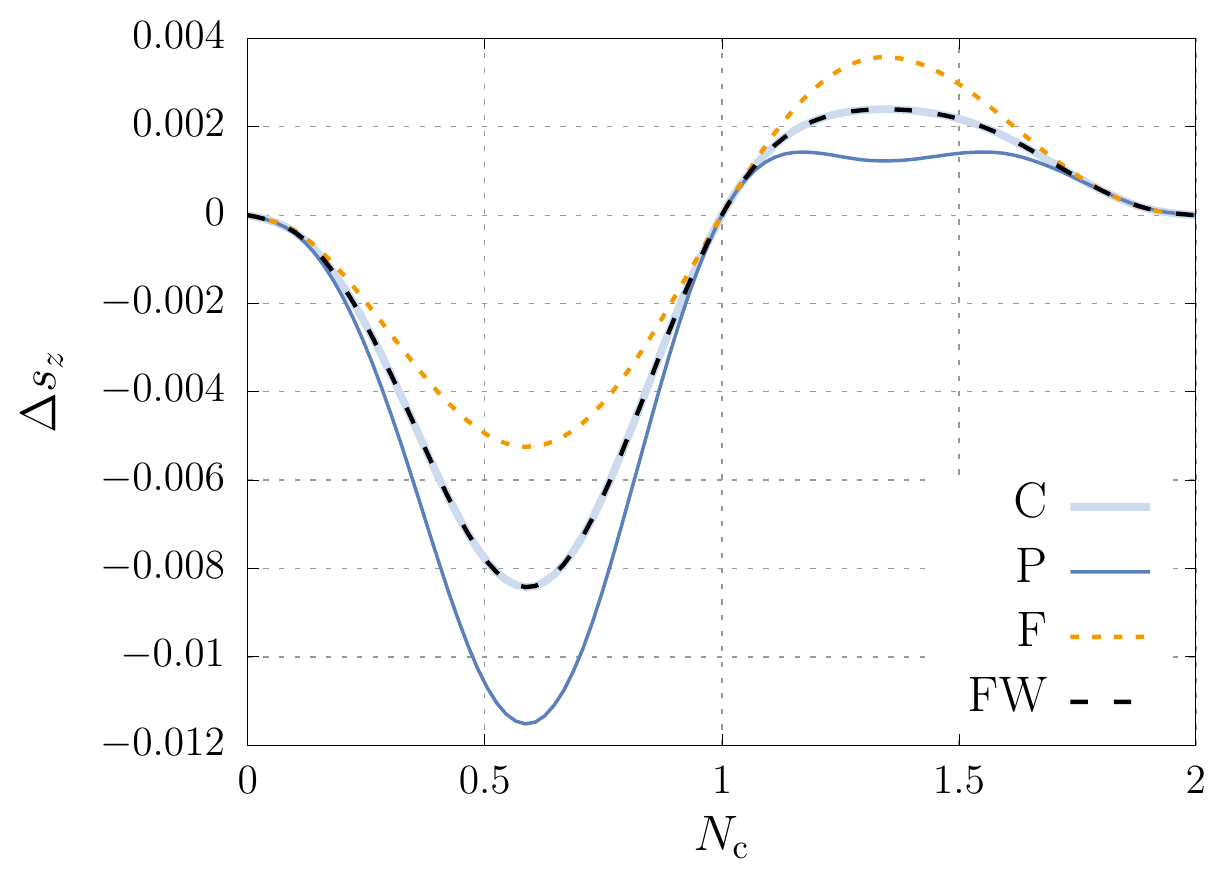}}
\caption{Change of the spin projections $s_x$~(left) and $s_z$~(right) of a relativistic electron wave packet after the interaction with the laser pulse~\eqref{eq:field_1}--\eqref{eq:field_2} as a function of $N_\text{c}$. The calculations are performed with the aid of the classical T-BMT equation~\eqref{eq:thomas} (line ``C'') and by solving the Dirac equation and using the Pauli, Frenkel, and Foldy-Wouthuysen spin operators (lines ``P'', ``F'', and ``FW'', respectively). The initial electron momentum is $p_z = 14$~a.u. ($p_x = p_y = 0$). The external field parameters are $E_* = 10$~a.u., $\omega = 1$~a.u.}
\label{fig:s_t_14}
\end{figure*}
\begin{figure*}
\subfigure{\label{fig:sx_t_70}\includegraphics[height=6.1cm]{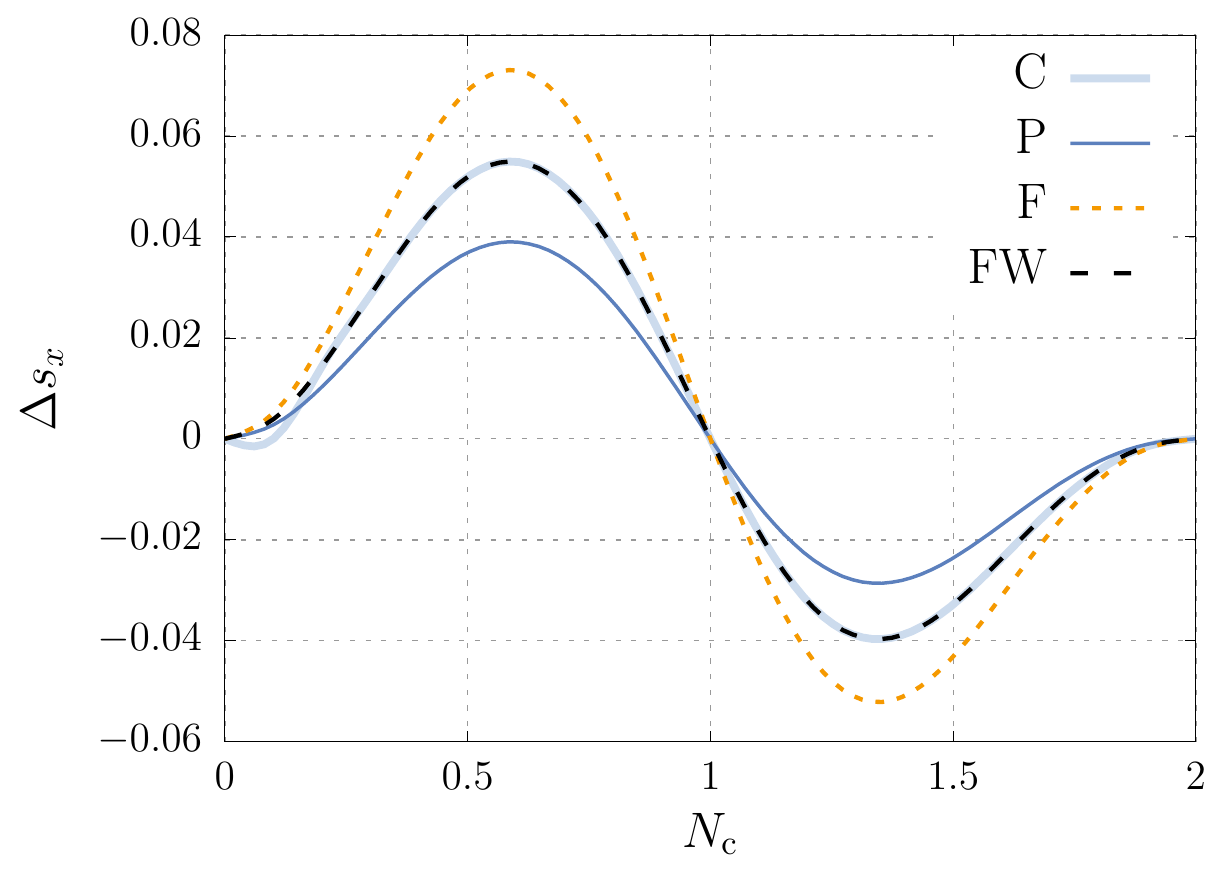}}~~~\subfigure{\label{fig:sz_t_70}\includegraphics[height=6.1cm]{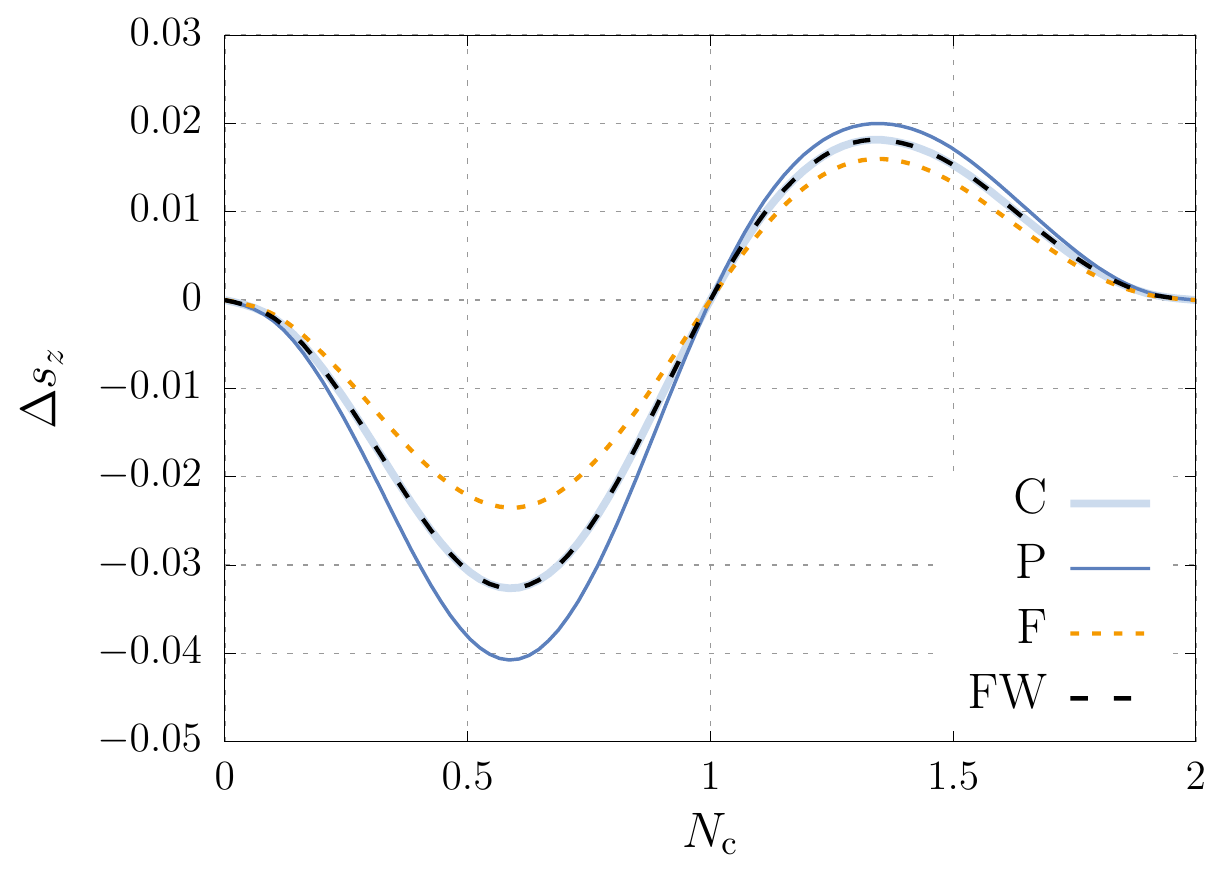}}
\caption{Change of the spin projections $s_x$~(left) and $s_z$~(right) of a relativistic electron wave packet after the interaction with the laser pulse~\eqref{eq:field_1}--\eqref{eq:field_2} as a function of $N_\text{c}$. The calculations are performed with the aid of the classical T-BMT equation~\eqref{eq:thomas} (line ``C'') and by solving the Dirac equation and using the Pauli, Frenkel, and Foldy-Wouthuysen spin operators (lines ``P'', ``F'', and ``FW'', respectively). The initial electron momentum is $p_z = 70$~a.u. ($p_x = p_y = 0$). The external field parameters are $E_* = 10$~a.u., $\omega = 1$~a.u.}
\label{fig:s_t_70}
\end{figure*}
\begin{figure*}[t]
\subfigure{\label{fig:sx_class_70_w01}\includegraphics[height=6.1cm]{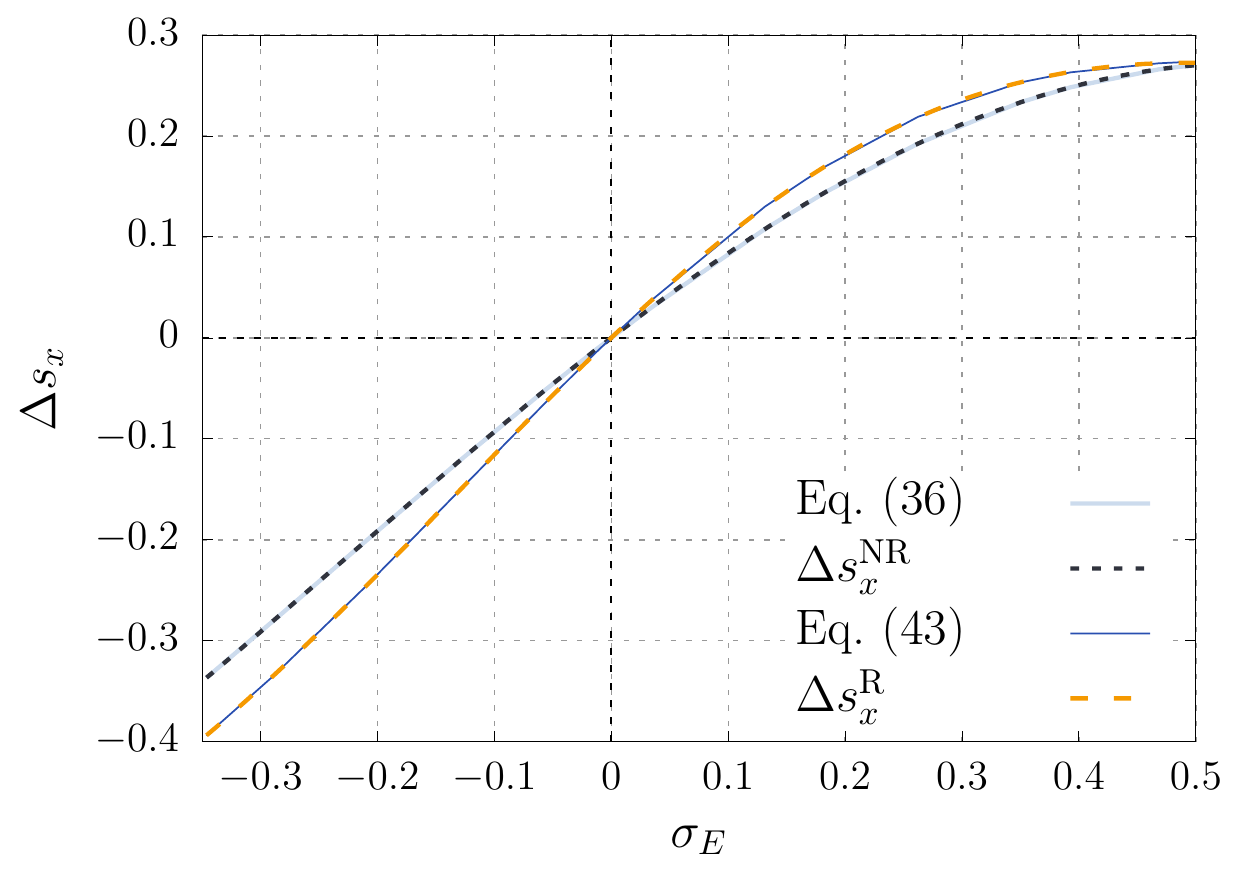}}~~~\subfigure{\label{fig:sz_class_70_w01}\includegraphics[height=6.1cm]{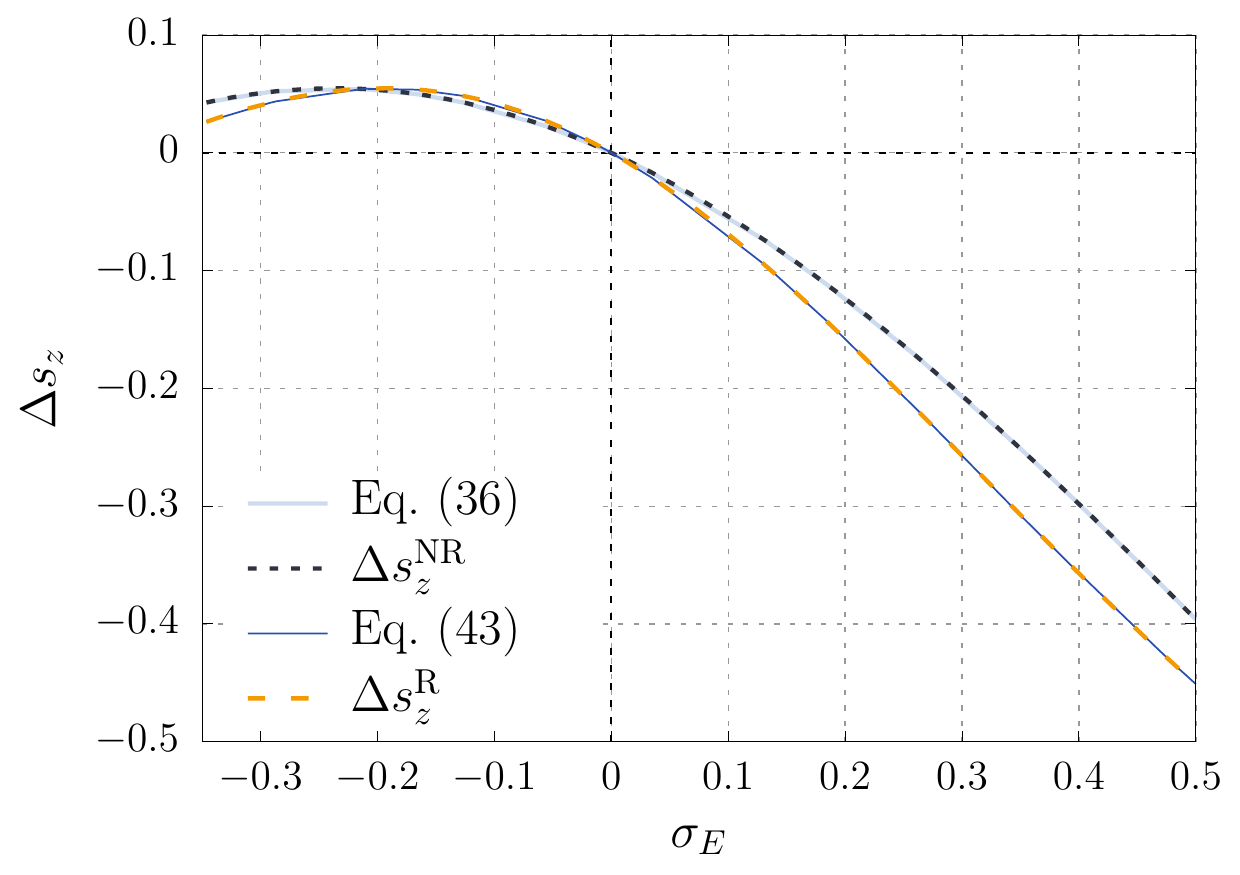}}
\caption{Change of the spin projections $s_x$~(left) and $s_z$~(right) of a classical electron after the interaction with the laser pulse~\eqref{eq:field_1}--\eqref{eq:field_2} as a function of the electric field area. The plot legend is the same as in Fig.~\ref{fig:s_class_14}. The initial momentum of the electron is $p_z=70$~a.u. ($p_x = p_y = 0$), and $\theta_0 = 0.472 \approx 27^\circ$. The external field parameters are $E_* = 10$~a.u., $\omega = 0.1$~a.u.}
\label{fig:s_class_70_w01}
\end{figure*}

\subsection{Quantum spin dynamics} \label{sec:res_quantum}

Using the method described in Sec.~\ref{sec:plane_wave}, we perform quantum computations of the electron spin projections as a function of $N_\text{c}$ for three different values of the initial electron momentum $p_z$. To calculate the mean values of the spin projections, we employ four different spin operators described in Sec.~\ref{sec:spin}: Pauli, Frenkel, Foldy-Wouthuysen, and Pryce ones (the Foldy-Wouthuysen and Pryce operators yield precisely the same data). The results will be compared with the classical predictions obtained by means of the T-BMT equation~\eqref{eq:thomas} being solved numerically, i.e., both quantum and classical treatment exactly take into account the finite-size effects.

As was stated above, the generalized momentum projections along the $x$ and $y$ directions are conserved since the external field depends only on the $z$ coordinate. It is useful to estimate the total change of the $z$ momentum component. Assuming that the main contribution in the expansion~\eqref{eq:expansion_psi_volkov} corresponds to the initial momentum $\boldsymbol{p}$ with $p_x = p_y = 0$ and taking into account that for $\xi \geqslant \xi_{\text{max}}$ the vector potential is constant, $\mathcal{A}(\xi) = \mathcal{A}_0 = c S_E$, we derive the total change $\Delta p_z$, which follows from Eq.~\eqref{eq:f_volkov}:
\begin{equation}
\Delta p_z \approx \frac{c}{\sqrt{c^2+p_z^2}-p_z} \frac{S_E^2}{2c}.
\label{eq:delta_pz}
\end{equation}
For $p_z/c \to 0$ it tends to $S_E^2/(2c)$.

Let us first consider the case $p_z=0$~(see Fig.~\ref{fig:s_t_0}). In Fig.~\ref{fig:s_t_0}(left) we observe that all of the spin operators give identical results that coincide with the classical curve. Moreover these lines exactly reproduce the curve in Fig.~\ref{fig:area} according to Eq.~\eqref{eq:dsx_nr_app_theta0} as the particle motion along the $z$ direction is essentially nonrelativistic (see discussion below). In Fig.~\ref{fig:s_t_0}(right) we observe a tremendously different situation. Using the Foldy-Wouthuysen operator predicts the same $N_\text{c}$ dependence as the classical T-BMT equation, whereas the Pauli and Frenkel operators lead to substantially different results. Since in the nonrelativistic limit all of these operators coincide, the discrepancy is associated with relativistic effects. In order to explain the difference in the behavior of the curves displayed in the two graphs, we shall consider the second term in the definition of the Frenkel operator~(\ref{eq:spin_F}), which corresponds to the leading relativistic correction to the Pauli operator [see also Eq.~\eqref{eq:spin_expansion}]. Since the field is polarized along the $x$ axis and we always assume $p_x = p_y = 0$, we receive
\begin{eqnarray}
(\hat{\boldsymbol{p}}+\boldsymbol{A}/c)\times \boldsymbol{\alpha} &=& -\hat{p}_z \alpha_y \boldsymbol{e}_x + [\hat{p}_z \alpha_x - (A_x/c)\alpha_z]\boldsymbol{e}_y \nonumber \\
{} &+& (A_x/c) \alpha_y \boldsymbol{e}_z. \label{eq:vector_product}
\end{eqnarray}
As the central value of the electron's initial momentum is zero ($p_z = 0$) and the $z$ component hardly changes in the laser field [$\Delta p_z \approx S_E^2/(2c) \lesssim 0.73$~a.u.], the $x$ projection of this vector product vanishes, which explains why we obtain the indistinguishable curves in Fig.~\ref{fig:s_t_0}(left). On the other hand, the relativistic dynamics of the $z$ component of the electron spin is much less trivial due to notable acceleration of the particle along the $x$ axis. The final value of the $x$ projection $A_x/c$ of the kinetic momentum is determined by the electric field area $S_E$, which reaches $\sim 14$~a.u. Note that for integer values of $N_\text{c}$, the area $S_E$ vanishes, so all the curves in Fig.~\ref{fig:s_t_0}(right) coincide and intersect with the line $\Delta s = 0$, while for large values of the pulse area~\eqref{eq:S_E_Nc}, the discrepancy is great. According to the results of our computations, the relativistic part of the Frenkel operator completely cancel the nonrelativistic (Pauli) contribution. Furthermore, the higher-order relativistic terms included in the Foldy-Wouthuysen operator make the quantum predictions exactly coincide with the classical estimates. 

In Fig.~\ref{fig:s_t_14}(left), a slight difference between the curves becomes noticeable due to the nonzero initial momentum $p_z = 14$~a.u. [see the $x$ projection of Eq.~\eqref{eq:vector_product}]. For the projection $s_z$, the relativistic effects, which arise not due to the nonzero value of $p_z$ but due to acceleration of the electron in the laser field, are clearly visible in Fig.~\ref{fig:s_t_14}(right). The use of the Foldy-Wouthuysen operator still leads to the classical results as it should be according to Refs.~\cite{Foldy1950, fradkin-good, silenko_2020}. Finally, in Fig.~\ref{fig:s_t_70} we depict the data obtained for the case $p_z =70$~a.u. The difference among various curves in the case of $\Delta s_x$ becomes evident. Nevertheless, the ``FW'' curve still accurately reproduces the classical predictions for both $\Delta s_x$ and $\Delta s_z$.

The results presented in Figs.~\ref{fig:s_t_0}--\ref{fig:s_t_70} bring us to two main conclusions. First, it was demonstrated that the classical predictions are reproduced to high precision by quantum simulations with the Foldy-Wouthuysen operator up to $p_z \approx c/2$. According to Sec.~\ref{sec:res_class}, it means that the relativistic electron spin dynamics is basically described by the classical approximate formulas~\eqref{eq:rel-spin_x}--\eqref{eq:rel-spin_z} and thus determined by the electric field area. The latter point indicates a great efficiency of unipolar laser pulses in the context of changing the electron spin state. Second, the spin dynamics is described in significantly different ways when using different spin operators. More specifically, the use of the Pauli and Frenkel operators does not lead to a quantitative coincidence of the results with the classical predictions, whereas the values obtained by employing the Foldy-Wouthuysen operator match the classical ones in all our calculations. These findings of numerical simulations confirm the correspondence between the classical spin vector and the quantum-mechanical spin operator in the form of Foldy-Wouthuysen~\cite{Foldy1950, fradkin-good, silenko_2020}.

So far the electric field area always obeyed $\sigma_E \lesssim 0.05$. In the next section, we will increase it by choosing a lower frequency. Accordingly, most of the relativistic effects discussed above as well as the changes of the spin projections themselves will become more pronounced. Namely, we will employ $\omega = 0.1$~a.u., so the field area will become ten times larger (the $N_\text{c}$ dependence in Fig.~\ref{fig:area} will be multiplied by $10$). This frequency relates to the wavelength $\lambda \approx 456$~nm, which corresponds to the visible spectrum.

\subsection{Lower frequency} \label{sec:res_low_freq}

First, we will consider the classical spin dynamics along the same lines as in Sec.~\ref{sec:res_class}. To keep the discussion concise, we present the results only for $p_z = 70$~a.u. (see Fig.~\ref{fig:s_class_70_w01}). As was stated above, the electric field area is now ten times larger, so the particle's spin changes much more significantly, and the simplest expressions~\eqref{eq:class-spin_x_nr} and \eqref{eq:class-spin_z_nr} are no longer applicable [for $\omega = 1$~a.u. they gave the same results as Eqs.~\eqref{eq:class-spin_x} and \eqref{eq:class-spin_z}].
\begin{figure*}
\subfigure{\label{fig:w01_sx_t_0}\includegraphics[height=6.1cm]{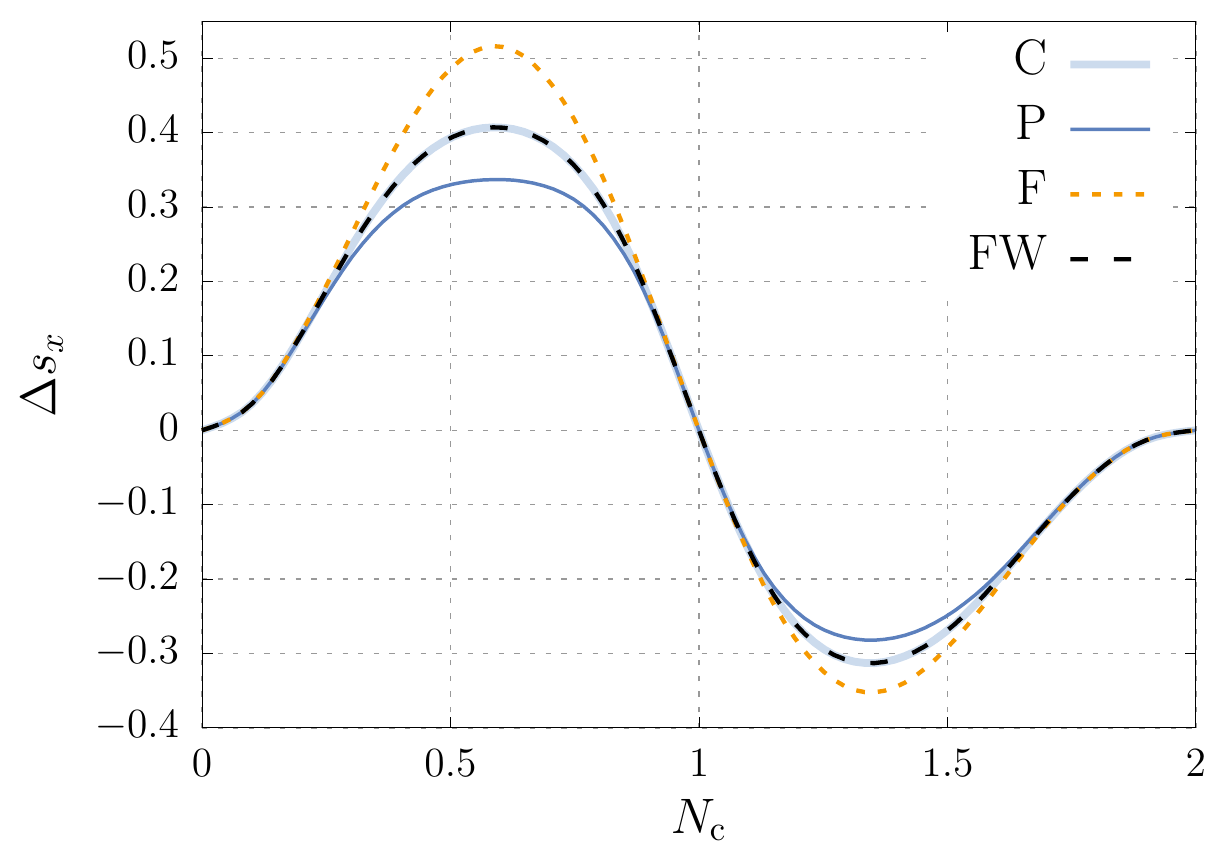}}~~~\subfigure{\label{fig:w01_sz_t_0}\includegraphics[height=6.1cm]{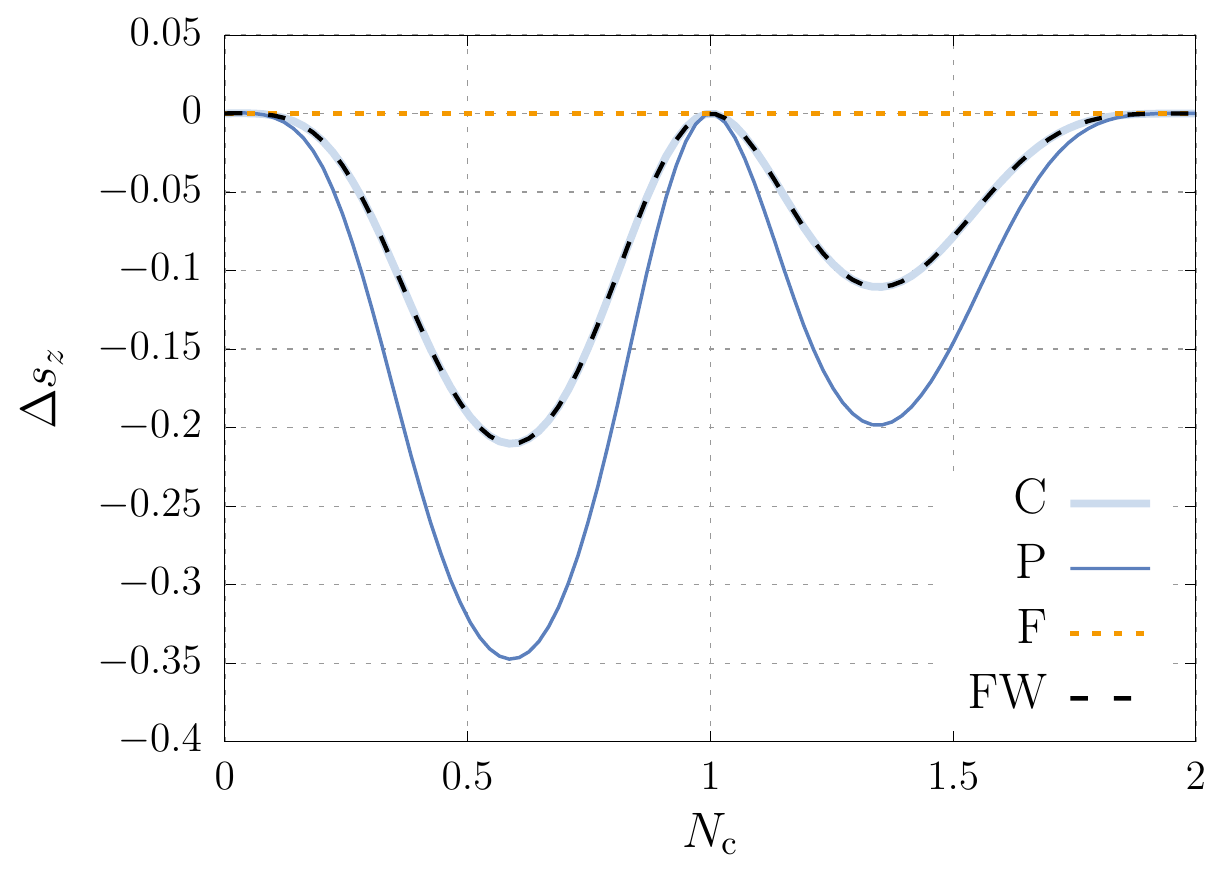}}
\caption{Change of the spin projections $s_x$~(left) and $s_z$~(right) of a relativistic electron wave packet after the interaction with the laser pulse~\eqref{eq:field_1}--\eqref{eq:field_2} as a function of $N_\text{c}$. The calculations are performed with the aid of the classical T-BMT equation~\eqref{eq:thomas} (line ``C'') and by solving the Dirac equation and using the Pauli, Frenkel, and Foldy-Wouthuysen spin operators (lines ``P'', ``F'', and ``FW'', respectively). The initial electron momentum is $\boldsymbol{p} = 0$. The external field parameters are $E_* = 10$~a.u., $\omega = 0.1$~a.u.}
\label{fig:w01_s_t_0}
\end{figure*}
\begin{figure*}
\subfigure{\label{fig:w01_sx_t_70}\includegraphics[height=6.1cm]{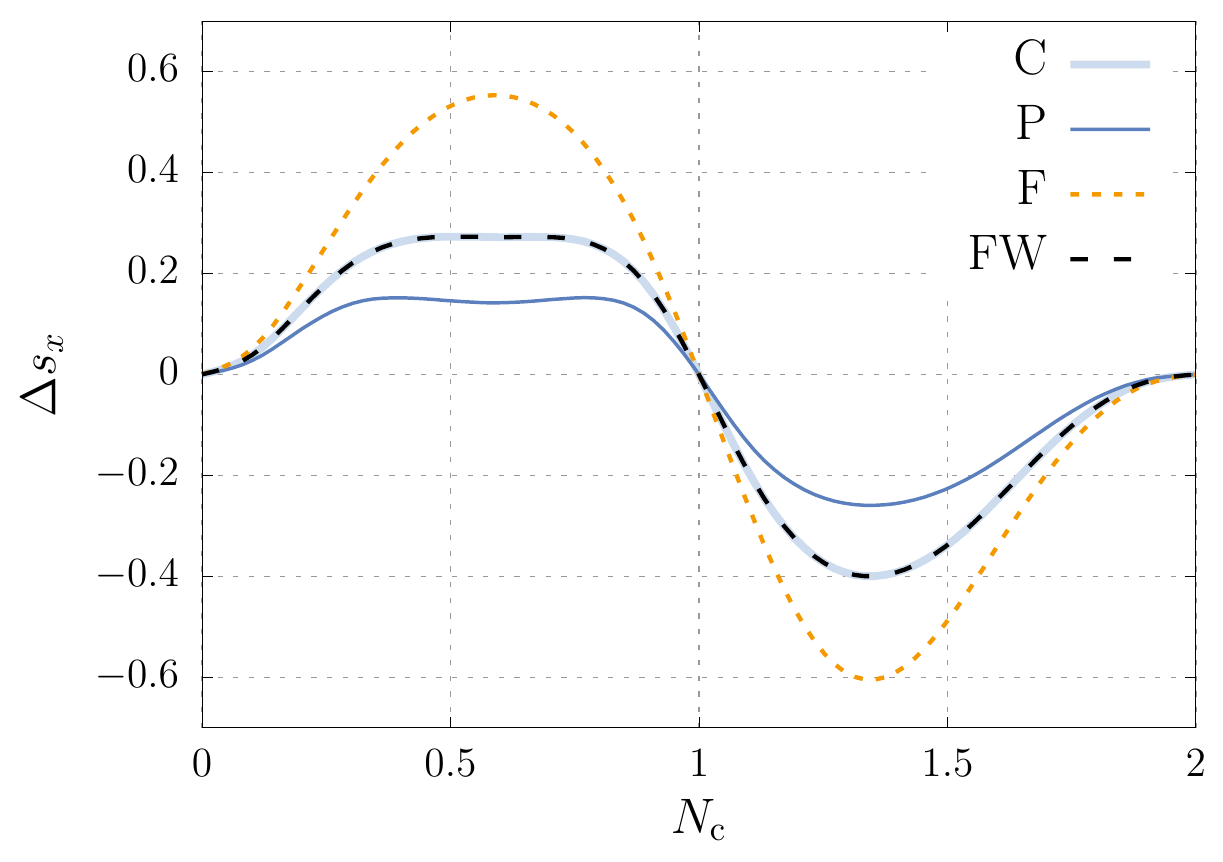}}~~~\subfigure{\label{fig:w01_sz_t_70}\includegraphics[height=6.1cm]{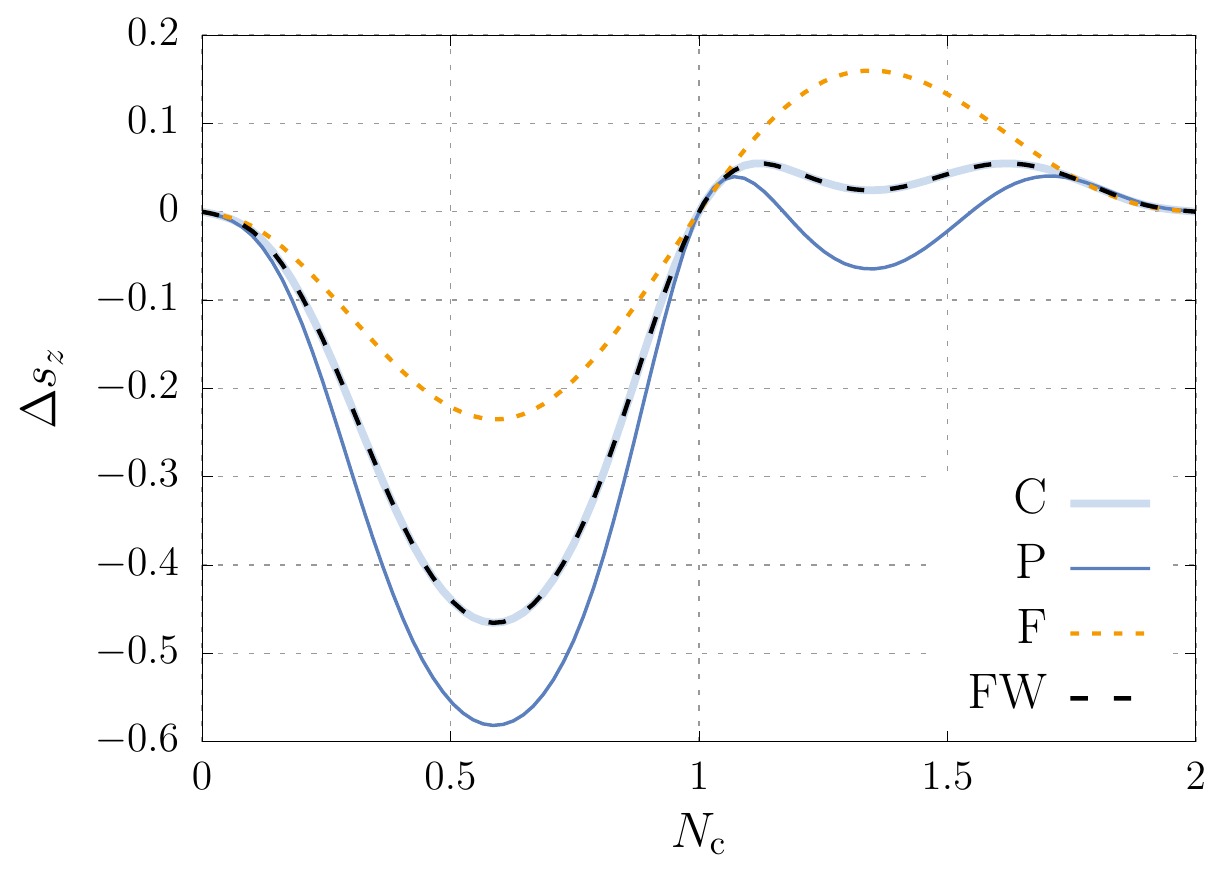}}
\caption{Change of the spin projections $s_x$~(left) and $s_z$~(right) of a relativistic electron wave packet after the interaction with the laser pulse~\eqref{eq:field_1}--\eqref{eq:field_2} as a function of $N_\text{c}$. The calculations are performed with the aid of the classical T-BMT equation~\eqref{eq:thomas} (line ``C'') and by solving the Dirac equation and using the Pauli, Frenkel, and Foldy-Wouthuysen spin operators (lines ``P'', ``F'', and ``FW'', respectively). The initial electron momentum is $p_z = 70$~a.u. ($p_x = p_y = 0$). The external field parameters are $E_* = 10$~a.u., $\omega = 0.1$~a.u.}
\label{fig:w01_s_t_70}
\end{figure*}

The main new feature here is the fact that the curves in Fig.~\ref{fig:s_class_70_w01} are far from being straight in contrast to those displayed in Fig.~\ref{fig:s_class_70}. The reason for this is the same --- the parameter $\sigma_E$ reaches too large values. Considering these plots in a sufficiently small vicinity of $\sigma_E = 0$, we would obtain the plots similar to those depicted in Fig.~\ref{fig:s_class_70}. Moreover, the curves in Fig.~\ref{fig:s_class_70_w01} are no longer parabolas as they should be described by means of the more complex expressions~\eqref{eq:class-spin_x}, \eqref{eq:class-spin_z} (NR) and~\eqref{eq:rel-spin_x} and \eqref{eq:rel-spin_z} (R), respectively. The most important point here is that these approximate closed-form expressions explicitly involving the electric field area remain very accurate, so there is still no need to perform the full computations based on Eqs.~\eqref{eq:larmor} and \eqref{eq:thomas}.

Finally, we turn to the quantum description of the process. In Fig.~\ref{fig:w01_s_t_0} we present the $N_\text{c}$ dependences for $\boldsymbol{p} = 0$ and $\omega = 0.1$~a.u. Even for $\Delta s_x$ the results are considerably different because the $z$ projection of the electron's momentum now notably changes. Both our quantum and classical computations confirmed Eq.~\eqref{eq:delta_pz} to high accuracy, i.e. $\Delta p_z \sim S_E^2/(2c)$ (this ratio is now 100 times greater and can reach $73$~a.u.). Nevertheless, the Foldy-Wouthhuysen operator leads to the same data as the classical calculations, whereas the other operators predict different patterns. Note also that when using the Frenkel operator, the $s_x$ projection can exceed $1/2$ (see dashed orange line in Fig.~\ref{fig:w01_s_t_0}) since the eigenvalues of this operator do not equal $\pm 1/2$.

In Fig.~\ref{fig:w01_s_t_70} we display our results for $p_z = 70$~a.u. which lead essentially to the same findings as those discussed above: the discrepancy among different curves is evident, the ``FW'' curve always coincides with the classical one. The initial value $p_z$ does not play now a decisive role as this momentum projection changes a lot under the action of the laser field.

For the parameters chosen in our computations, we did not observe any significant difference between the results of quantum calculations and those obtained by means of the T-BMT equation. According to the common criteria justifying a quasiclassical treatment~(see, e.g., Ref.~\cite{landau3} for the nonrelativistic conditions), some discrepancy may appear in the domain of small particle's momenta or high laser frequencies. However, this regime corresponds to a smaller field area obscuring the spin effects, which we are interested in.

\section{Conclusion}\label{sec:conclusion}

In this work, we analyzed the dynamics of the electron spin in the field of a linearly polarized short laser pulse of a finite size. First, it was demonstrated that the total change of the classical spin can be described by simple closed-form expressions involving the initial momentum of the particle and the electric field area of the laser pulse. Our quantum computations based on the Dirac equation indicated also that the pulse area is paramount within the process under consideration. In order to maximize the impact that the laser field has on the electron spin, one has to generate pulses with a larger electric field area.

Second, unipolar pulses may allow one to directly probe the relativistic spin operators and to assess their relevance to the observable quantities. It was shown that the different choice of the relativistic spin operator can indeed lead to significantly different results, and the corresponding discrepancies strongly depend on the field area. In particular, it turned out that the predictions obtained by using the Foldy-Wouthuysen spin operator always match the classical results. This point confirms that the Foldy-Wouthuysen operator is the quantum-mechanical counterpart of the classical spin. Moreover, since the initial and final states of the electron are free, the Pryce operator yields the same results as that of Foldy-Wouthuysen. The other forms of the spin operator (Pauli and Frenkel ones) predict substantially different patterns. Besides, instead of using the Foldy-Wouthuysen operator, one can equivalently perform the Lorentz boost to the particle's rest frame and calculate the mean values of the Pauli spin operator since the wave packet does not contain any contributions from the negative energy continuum.

\begin{acknowledgments}
This work was supported by Russian Foundation for Basic Research (RFBR) (Grant No.~19-02-00312). The calculations were performed at the Computing Center of Saint Petersburg State University Research Park.
\end{acknowledgments}

\appendix*

\section{Exact solution of the T-BMT equation in the case of a monochromatic plane wave}

Here we present a derivation of the exact solution of the classical equation~\eqref{eq:thomas} governing the spin dynamics in the case of a nonzero initial momentum $p_z$ (for $p_z=0$ it can be found in Ref.~\cite{Walser_2002}). The external field is assumed to be a monochromatic plane wave~\eqref{eq:field_mono}.

First, one has to solve the relativistic equations of motion for a classical electron. They read
\begin{eqnarray}
m \frac{d\boldsymbol{u}}{dt} &=& e \Big ( \boldsymbol{E} + \frac{\boldsymbol{v}}{c} \times \boldsymbol{B} \Big ), \label{eq:app:eq_mot_momentum}\\
\frac{d\varepsilon}{dt} &=& e \boldsymbol{v} \boldsymbol{E},
\label{eq:app:eq_mot_energy}
\end{eqnarray}
where $\boldsymbol{u} = \gamma \boldsymbol{v}$, $\varepsilon = \gamma mc^2$, $\gamma = (1-v^2/c^2)^{-1/2}$, and we have recovered the electron charge $e$ and mass $m$. The $y$~component of Eq.~\eqref{eq:app:eq_mot_momentum} leads to $u_y (t) = v_y (t) = 0$ as the initial conditions are $u_x(0) = u_y(0) = 0$, $u_z(0) = p_z/m$. Here $p_z$ represents a specific value of the initial momentum projection (unlike $\boldsymbol{u}$, $\boldsymbol{v}$, $\gamma$, and $\varepsilon$, it does not depend on time). The $x$~component of Eq.~\eqref{eq:app:eq_mot_momentum} has the following form:
\begin{equation}
m \frac{du_x}{dt} = e \Big ( 1 - \frac{v_z}{c} \Big ) E_* \cos (\omega t - kz).
\label{eq:app:ux}
\end{equation}
It is convenient to substitute $t$ with $\tau \equiv t - z/c$. Since $d\tau = (1 - v_z/c) \, dt$, one obtains
\begin{equation}
m \frac{du_x}{d\tau} = e E_* \cos \omega \tau,
\label{eq:app:ux_tau}
\end{equation}
and thus
\begin{equation}
u_x (\tau) = u_0 \sin \omega \tau,
\label{eq:app:ux_tau_sol}
\end{equation}
where $u_0 \equiv eE_*/(m\omega)$. The particle is initially at $z=0$, which means that $\tau = 0$ is equivalent to $t=0$. Using then the $z$~component of Eq.~\eqref{eq:app:eq_mot_momentum} and Eq.~\eqref{eq:app:eq_mot_energy}, one can easily obtain $d\varepsilon = mc \, du_z$, which brings us to
\begin{equation}
\varepsilon = mc^2 \sqrt{1 + \frac{p_z^2}{(m c)^2}} - c p_z + mc u_z.
\label{eq:app:energy_uz}
\end{equation}
The differential equation for $u_z (\tau)$, which follows either from Eq.~\eqref{eq:app:eq_mot_momentum} or from Eq.~\eqref{eq:app:eq_mot_energy}, reads
\begin{equation}
m \frac{du_z}{d\tau} = \frac{u_x}{\gamma c - u_z} \, eE_* \cos \omega \tau.
\label{eq:app:uz_ux}
\end{equation}
Using then Eqs.~\eqref{eq:app:ux_tau_sol}--\eqref{eq:app:uz_ux} and $\gamma = \varepsilon/(mc^2)$, we receive
\begin{equation}
\frac{du_z}{d\tau} = \frac{\omega u_0^2}{2c} \, \frac{\sin 2\omega \tau}{\sqrt{1 + [p_z/(m c)]^2} - p_z/(mc)}.
\label{eq:app:uz_de}
\end{equation}
Integrating this equation and taking into account $u_z (0) = p_z/m$, we obtain
\begin{equation}
u_z(\tau) = \frac{p_z}{m} + \frac{u_0^2}{2c} \, \frac{\sin^2 \omega \tau}{\sqrt{1 + [p_z/(m c)]^2} - p_z/(mc)}.
\label{eq:app:uz_sol}
\end{equation}

Let us now discuss how the T-BMT equation~\eqref{eq:thomas} can be solved. First, we note that the factor $-1/c$ in Eq.~\eqref{eq:thomas} corresponds to $e/(mc)$. Second, from Eq.~\eqref{eq:thomas} it immediately follows that $s_y=\text{const}$. The equations involving $s_x$ and $s_z$ have the following form:
\begin{eqnarray}
\frac{ds_x}{dt} &=& -\frac{\omega u_0}{\gamma c} \, s_z \bigg (1 - \frac{u_z/c}{\gamma + 1} \bigg ) \cos (\omega t - kz), \label{eq:app:sx_eq}\\
\frac{ds_z}{dt} &=& \frac{\omega u_0}{\gamma c} \, s_x \bigg (1 - \frac{u_z/c}{\gamma + 1} \bigg ) \cos (\omega t - kz). \label{eq:app:sz_eq}
\end{eqnarray}
Using now the ansatz $s_x = (1/2) \sin \theta$ and $s_z = (1/2) \cos \theta$, we derive a differential equation for $\theta (\tau)$:
\begin{equation}
\frac{d\theta}{d\tau} = -\frac{\omega u_0}{c} \, \frac{\gamma + 1 - u_z/c}{(\gamma + 1)(\gamma - u_z/c)} \, \cos \omega \tau.
\label{eq:app:theta_eq}
\end{equation}
Having obtained the functions $u_z (\tau)$ and $\gamma (\tau) = \varepsilon (\tau) / (mc^2)$ [see Eqs.~\eqref{eq:app:energy_uz} and \eqref{eq:app:uz_sol}], one can now find
\begin{eqnarray}
\theta (\tau)&& = \theta_0 -  2\arctan \bigg\{\frac{1}{D}\frac{u_0}{2c}\sin{\omega\tau} \bigg\}\label{eq:app:theta_sol_speed}\\
&&=\theta_0 +  2\arctan \bigg\{\frac{1}{D} \sigma_E (\tau) \bigg\},
\label{eq:app:theta_sol}
\end{eqnarray}
where $D$ and $\sigma_E(\tau)$ are defined in the same way as in Sec.~\ref{sec:class}:
\begin{eqnarray}
D &=& \frac{1}{2}\bigg[1 + \Pi_z - \frac{p_z}{mc}\bigg],\\
\Pi_z &=& \sqrt{1+\frac{p^2_z}{(mc)^2}}, \\
\sigma_E(\tau) &=& \frac{|e| S_E (\tau)}{2mc}.
\end{eqnarray}
It is worth noting that $\theta (\tau)$ expressed in terms of the electric field area of the pulse [see Eq.~\eqref{eq:app:theta_sol}] does not depend on the initial phase of the field. Indeed, if we add some initial phase $\varphi_0$ to the argument of cosine in~\eqref{eq:field_mono}, the expression~\eqref{eq:app:ux_tau_sol} will turn into
\begin{equation}
u_x (\tau) = u_0 \big[\sin{(\omega\tau+\varphi_0)} -\sin{\varphi_0}\big],
\label{eq:app:ux_tau_sol_phi0}
\end{equation}
and the expression~\eqref{eq:app:theta_sol_speed} will be modified accordingly. However, due to the fact that in this case the electric field area has the form
\begin{equation}
    S_E(\tau)=\frac{E_{*}}{\omega}\big[\sin{(\omega\tau+\varphi_0)} -\sin{\varphi_0}\big],
\end{equation}
the connection between $\sigma_E(\tau)$ and $u_x(\tau)$ remains unchanged,
\begin{eqnarray}
   \sigma_E(\tau)= - \frac{u_x (\tau)}{2c}.
\end{eqnarray}
Hence, we can conclude that regardless the presence of $\varphi_0$ the function $\theta(\tau)$ is fully determined by $\sigma_E(\tau)$.

Thus, the spin projections in terms of $\theta(\tau)$ change according to the following relations:
\begin{eqnarray}
\Delta s_x (\tau) &=& \frac{1}{2} \big [ \sin \theta (\tau) - \sin \theta_0 \big ]\nonumber\\
&=& \sin\bigg[\arctan\bigg\{\frac{\sigma_E (\tau)}{D}\bigg\}\bigg]\nonumber\\
&\times&\cos\bigg[\theta_0+\arctan\bigg\{\frac{\sigma_E (\tau)}{D}\bigg\}\bigg], \\
\Delta s_z (\tau) &=& \frac{1}{2} \big [ \cos \theta (\tau) - \cos \theta_0 \big ]\nonumber\\
&=& - \sin\bigg[\arctan\bigg\{\frac{\sigma_E (\tau)}{D}\bigg\}\bigg]\nonumber\\
&\times& \sin\bigg[\theta_0+\arctan\bigg\{\frac{\sigma_E (\tau)}{D}\bigg\}\bigg],
\end{eqnarray}
which coincide with those discussed in the main text [see Eqs.~\eqref{eq:rel-spin_x} and \eqref{eq:rel-spin_z} and comments below them].

\bibliography{lit}

\end{document}